\documentclass[a4paper,11pt]{article}
\pdfoutput=1 

\usepackage{jcappub} 

\usepackage[T1]{fontenc} 

\title{Constraining bursty star formation histories with galaxy UV and H$\alpha$ luminosity functions and clustering}

\author[a,1]{Guochao Sun,\note{Corresponding author.}}
\author[b]{Julian B. Mu\~{n}oz,}
\author[c,d]{Jordan Mirocha}
\author[a]{and Claude-Andr\'{e} Faucher-Gigu\`{e}re}

\affiliation[a]{CIERA and Department of Physics and Astronomy, Northwestern University,\\1800 Sherman Ave, Evanston, IL 60201, USA}
\affiliation[b]{Department of Astronomy, The University of Texas at Austin,\\2515 Speedway, Stop C1400, Austin, TX 78712, USA}
\affiliation[c]{Jet Propulsion Laboratory, California Institute of Technology, 4800 Oak Grove Drive, Pasadena, CA 91109, USA}
\affiliation[d]{California Institute of Technology,  1200 E. California Boulevard, Pasadena, CA 91125, USA}

\emailAdd{guochao.sun@northwestern.edu}
\emailAdd{julianbmunoz@utexas.edu}
\emailAdd{jordan.mirocha@jpl.nasa.gov}
\emailAdd{cgiguere@northwestern.edu}

\abstract{The observed prevalence of galaxies exhibiting bursty star formation histories (SFHs) at $z\gtrsim6$ has created new challenges and opportunities for understanding their formation pathways. The degenerate effects of the efficiency and burstiness of star formation on the observed UV luminosity function are separable by galaxy clustering. However, quantifying the timescales of burstiness requires more than just the continuum UV measurements. Here we develop a flexible semi-analytic framework for modeling both the amplitude of star formation rate (SFR) variations and their temporal correlation, from which the luminosity function and clustering can be derived for SFR indicators tracing different characteristic timescales (e.g., UV continuum and H$\alpha$ luminosities). Based on this framework, we study the prospect of using galaxy summary statistics to distinguish models where SFR fluctuations are prescribed by different power spectral density (PSD) forms. Using the Fisher matrix approach, we forecast the constraints on parameters in our PSD-based model that can be extracted from mock JWST observations of the UV and H$\alpha$ luminosity functions and clustering bias factors at $z\sim6$. If potential confusion due to e.g., dust attenuation and stellar population effects can be properly quantified, these results imply the possibility of probing the burstiness of high-$z$ galaxies with one-point and two-point statistics and highlight the benefits of combining long-term and short-term SFR tracers. Our flexible framework can be readily extended to characterize the SFH of high-redshift galaxies with a wider range of observational diagnostics.}

\keywords{galaxy formation, high redshift galaxies, semi-analytic modeling, star formation}

\begin{document}
\maketitle
\flushbottom

\section{Introduction}
\label{sec:intro}

The advent of the James Webb Space Telescope (JWST) has not only enabled galaxies to be routinely discovered well into the Cosmic Dawn era ($z>6$) but also revealed an unprecedented amount of information about how these early galaxies assembled their stars and evolved. The dominant presence of galaxies with bursty star formation histories (SFHs) is one of the most intriguing aspects of early galaxy formation discovered by recent JWST observations \cite{Ciesla2024,Dressler2024,Endsley2024}. Spectral energy distribution (SED) modeling of galaxy photometry implies that a substantial fraction of galaxies at $z \gtrsim 6$ have SFHs with one or multiple peaks and thus strong time variability, which are further supported by measurements of star formation rate (SFR) indicators (e.g., UV continuum and Balmer line luminosities) sensitive to variations of the SFR on different timescales \cite{Asada2024}. Meanwhile, the spectroscopic discovery of low-mass, quiescent galaxies in the same epoch provides further evidence for highly stochastic SFHs, which allow low-mass galaxies to temporarily but rapidly quench by stellar feedback following starbursts \cite{Gelli2023,Dome2024,FaisstMorishita2024,Looser2024}. These highly bursty SFHs are in stark contrast with the smooth, continuous SFHs inferred for typical, more massive star-forming galaxies at lower redshifts such as the Milky Way, but resemble those of local dwarf galaxies, for which similar signs of bursty SFHs are often seen \cite{Weisz2012,Emami2019}. Understanding the physical drivers and observational implications of bursty star formation is therefore essential for building a complete picture of galaxy formation, especially in the high-$z$ universe. 

Constraints on the SFH of individual galaxies are useful for examining the physics of the SFR variability but come with many challenges. Each galaxy is essentially observed only at one snapshot in its evolutionary history. While physical details of the stellar populations formed at different times are encoded in the galaxy SED, which can be forward modeled by stellar population synthesis (SPS), reliably obtaining the SFH is a non-trivial task that demands high-quality spectroscopic or multi-band photometric data. Moreover, to reconstruct the SFH one has to deal with many sources of degeneracy, including physical properties of the source population such as the initial mass function (IMF), metallicity, and binarity, the presence of nebular emission, as well as complicating factors associated with dust attenuation. While modern SED-fitting tools have allowed the SFH to be highly flexible based on either parametric or non-parametric assumptions \cite{Carnall2018,Boquien2019,Johnson2021,Cappellari2023}, SFHs derived for individual galaxies are often uncertain and model dependent, especially when high S/N spectra are lacking. 

Alternative to reconstructing the entire SFH, it is possible and sometimes preferable to quantify the SFH burstiness, namely the short-term ($\lesssim 10\,$Myr) time variability of the current SFR \cite{Broussard2019,CT2019}, directly. This is most commonly done by contrasting SFR indicators sensitive to SFR variations on different timescales. For example, hydrogen Balmer lines like H$\alpha$ and H$\beta$ are recombinations in HII regions ionized by short-lived O/B stars and thus sensitive to SFR changes over as short as a few Myr, whereas the UV continuum is contributed also by longer-lived A stars and thus sensitive to changes over much longer timescales ($\sim$10--100\,Myr depending on the SFH) \cite{Weisz2012,Dominguez2015,Sparre2017,Emami2019,Faisst2019,JFV2021,Atek2022,Tacchella2022,Sun2023,Asada2024,Clarke2024}. By comparing the scatters of SFR values inferred from these indicators respectively or examining their joint distribution (e.g., the luminosity ratio $L_\mathrm{H\alpha}/\nu L_{\nu,1500}$), one can quantify the level of burstiness in the SFH of different galaxy populations selected by physical properties such as mass and redshift. While past studies have demonstrated the power of these simple statistics of SFR indicators for quantifying burstiness, such analyses mainly focus on specific samples of galaxies. It remains to be understood more generally how summary statistics of the entire galaxy population may be sensitive to bursty SFHs. This is of particular interest since the galaxy-halo connection and the light-to-mass ratio of galaxies depend significantly on the burstiness of star formation \cite{Carnall2019,Alarcon2023}, which makes it challenging to construct clean mass-selected samples from observations of galaxies with strongly bursty SFHs. Summary statistics like the galaxy luminosity function (LF) and clustering that characterize the one-point and two-point correlation of galaxies, respectively, are thus potentially useful measures of burstiness at the population level. 

Recently, bursty star formation and a high efficiency of converting baryons to stars have been proposed as two viable explanations for the galaxy UV LF at $z \gtrsim 10$ measured by the JWST \cite{Dekel2023,Harikane2023,Mason2023,Mirocha2023,Shen2023,Sun2023b,KB2024,Nikolic2024}. While an elevated bright-end UVLF can be produced in both cases, the degenerate effects of them have been shown to be separable by the (effective) bias factor measurable from the galaxy clustering \cite{Munoz2023,ChakrabortyChoudhury2024,Gelli2024}. This has provided the concrete evidence that these summary (one-point and two-point) statistics of galaxy number counts can be used to constrain SFHs on $\sim 100$ Myr timescales. A natural question then is to what level of detail bursty SFHs may be studied with galaxy summary statistics. In \cite{Munoz2023}, the authors made the first attempt to address this question by demonstrating that a constant scatter in the UV magnitude of galaxies, which relates to bursty SFHs encapsulated by a fixed scatter in the logarithmic SFR, can be unambiguously constrained by measuring the clustering bias factor as a function of the UV magnitude. This simple parameterization aims to capture the signature of bursty SFHs on the width of the UV magnitude distribution, but it does not come with a description of bursty SFHs themselves and only the longer term SFR variability is probed by the UV magnitude. As a result, it contains limited amount information about physical drivers of the burstiness and cannot be easily generalized to predict other spectral features and observables of interest. 

In this paper, motivated by the aforementioned probes of bursty SFHs based on joint H$\alpha$ and UV observations and the galaxy summary statistics, we extend the concept introduced in \cite{Munoz2023} by developing a flexible semi-analytic framework for forward modeling both the amplitude of SFR variations and their temporal correlation. The SFR fluctuations are described as a one-dimensional (1D) gaussian random field, which in turn can be specified by a parametric model of the power spectral density (PSD) \cite{CT2019,Iyer2020,Tacchella2020,PK2023,KB2024} or periodogram \cite{PF2023,Dome2025}. We combine this model with the SPS to estimate the UV- and H$\alpha$-based galaxy LFs and clustering bias factors under model assumptions corresponding to different scenarios of bursty star formation. Using a Fisher matrix analysis, we then demonstrate the information gain from combining the summary statistics of both UV and H$\alpha$ data to constrain bursty SFHs, especially their temporal structure, from mock JWST observations of the LF and clustering bias at $z = 6$. 

The remainder of this paper is organized as follows. In section~\ref{sec:models}, we describe our modeling framework for the bursty SFH that is specified by a 3-parameter PSD model, the SFR indicators and their correspondent summary statistics, and the Fisher-matrix formalism for estimating the parameter constraints from mock observations. In section~\ref{sec:results}, we use results of two reference PSD models to demonstrate how different scenarios of bursty star formation can be constrained and distinguished by the UV and H$\alpha$ LFs and bias factors. We discuss limitations of the current framework and outline a number of possible extensions in section~\ref{sec:discussion}, before concluding in section~\ref{sec:conclusions}. Throughout, we assume a flat, $\Lambda$CDM cosmology consistent with measurements by Planck \cite{Planck2016}. 

\section{Models}
\label{sec:models}

\subsection{Star formation histories} \label{sec:models:sfh}

To model the SFH with a prescribed, variable amount of burstiness, we decompose the SFH into two components: a smooth one that only slowly evolves over time at a given halo mass and a bursty component that describes potentially strong SFR fluctuations around the smooth one on shorter timescales. The SFH is then the sum of the two components, each as a time series of the SFR, namely
\begin{equation}
\log \mathrm{SFR}(t) = \log \langle \mathrm{SFR} \rangle(t) + \Delta \log \mathrm{SFR}(t).
\label{eq:sfh}
\end{equation}
Note that the SFH can be a function of not only cosmic time but also other physical quantities, such as the galaxy mass and environment. Throughout, we assume that the mass dependence is entirely carried by the smooth component in eq. (\ref{eq:sfh}), whereas the bursty component only sets the amplitude and temporal correlation of the logarithmic SFR that are mass-independent and time-invariant. In reality, this is likely an oversimplification since the level of burstiness in the SFH can have non-trivial dependence on galaxy physical properties like mass, redshift, and environment \cite{CAFG2018,FM2022,Gelli2024,KB2024,MenonPower2024}. However, to keep our models simple, we neglect this effect in this proof-of-the-concept study. Caveats and ways to account for such mass dependence will be discussed in section~\ref{sec:discussion}. 

There are several motivations for such a decomposition. First, in the high-redshift regime, host halos of galaxies grow exponentially fast, making the long-term evolution of a galaxy's SFH tightly correlated with the mass assembly history of its host halo \cite{Tacchella2018,WechslerTinker2018}. On the other hand, short-term variations in the SFH are typically associated with a broader set of physical drivers, including not only sources of scatter in the halo mass accretion history, e.g., mergers and local environments \cite{MihosHernquist1994,McBride2009,Kelson2016}, but also physics on smaller scales, e.g., stellar feedback and the cycling of gas through the halo \cite{Angles-Alcazar2017,Sparre2017} . Meanwhile, the isolation of a purely fluctuating component makes it convenient to prescribe the short-term time variability of the SFH as a gaussian random field, which can be easily parameterized. Meanwhile, this decomposition has been recently considered in parametric studies of the SFH \citep{KB2024,PK2023,Wan2024} and is supported by the analysis of SFHs extracted from high-resolution cosmological hydrodynamical simulations \cite{PF2023}. Next, we will explain how these two components can be modeled separately and justify our model assumptions. 

\subsubsection{A smooth component based on abundance matching}

The way we model the smooth component of the SFH, $\langle \mathrm{SFR} \rangle (t)$, formally resembles that considered in many semi-empirical studies of high-redshift galaxy formation \cite{SF2016,Mirocha2017,Tacchella2018,SL2024}. Specifically, we follow \cite{Mirocha2017} and assume that the mean SFR of galaxies traces the mass accretion rate onto dark matter halos by a redshift-independent star formation efficiency (SFE), $f_{\star}$, that can be parameterized as a double-power law in mass, namely
\begin{equation}
\langle \mathrm{SFR} \rangle (t) = f_{\star}(M_\mathrm{h})(\Omega_\mathrm{b}/\Omega_\mathrm{m}) \dot{M}_\mathrm{h}(t)
\end{equation}
and
\begin{equation}
f_{\star}(M_\mathrm{h}) = \frac{2 f_{\star,0}}{\left(\frac{M_\mathrm{h}}{M_\mathrm{p}}\right)^{\gamma_\mathrm{lo}} + \left(\frac{M_\mathrm{h}}{M_\mathrm{p}}\right)^{\gamma_\mathrm{hi}}}. 
\label{eq:sfe}
\end{equation}
Values of parameters $f_{\star,0}$, $M_\mathrm{p}$, $\gamma_\mathrm{lo}$, and $\gamma_\mathrm{hi}$ can be found such that the implied galaxy UV LF, stellar mass function, and so forth match the observations. Here, we take $f_{\star,0}=0.015$, $M_\mathrm{p}=3.6\times10^{11}\,M_{\odot}$, $\gamma_\mathrm{lo}=-0.55$, and $\gamma_\mathrm{hi}=0.77$ following \cite{Sun2021}. The halo mass accretion rate $\dot{M}_\mathrm{h}$ is determined from abundance-matching the halo mass function across cosmic times \cite{Furlanetto2017}, which guarantees the agreement between the mass and redshift dependence of $\dot{M}_\mathrm{h}$ and the evolution of the halo mass function assumed. While simple, this parameterization is a useful starting point for describing the galaxy--halo connection at high redshift and by construction reproduces the abundance of galaxies as a key observational constraint. More sophisticated models allowing for e.g., the redshift evolution of $f_{\star}(M_\mathrm{h})$ can be similarly constructed, though we note that current observations are still compatible with a non-evolving (but mass-dependent) SFE \cite{Feldmann2024}. 

\begin{figure}[htbp]
\centering
\includegraphics[width=\textwidth]{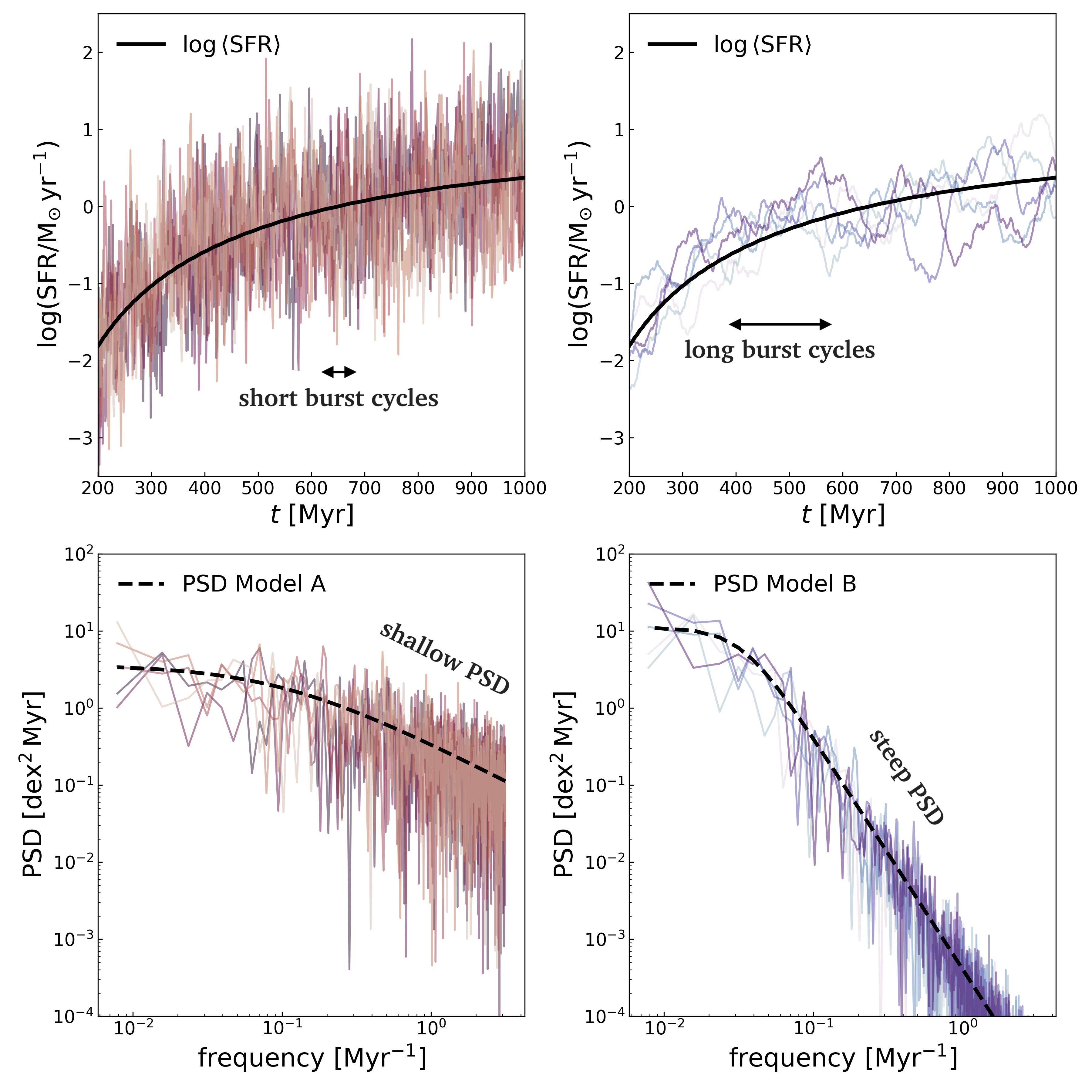}
\caption{Our two models for bursty SFHs (top) and the corresponding PSD (bottom) specified by different forms of the broken power-law model defined in eq. (\ref{eq:psd}), which are chosen to have the same magnitude of long-term $\log\mathrm{SFR}$ variations but differ in the short-term variability. In each column, 5 example realizations of the net SFH (including both smooth and bursty components) and their derived PSDs are plotted in color, whereas the input smooth SFH and PSD model are plotted in black. The correspondence between the PSD shape and the short-term SFH variability is clearly visible from the comparison of top and bottom panels.}
\label{fig:sfh-psd}
\end{figure} 

\subsubsection{A bursty component parameterized by the PSD}
Defining $\eta(t) = \Delta \log \mathrm{SFR}(t)$, the fluctuating component of the SFH (see figure~\ref{fig:sfh-psd}), which is assumed to be a gaussian random variable as a function of cosmic time $t$, we can describe the temporal correlation of $\eta$ by
\begin{equation}
\langle \eta(t) \eta(t') \rangle = \xi_{\eta}(|t - t'|), 
\end{equation}
where the auto-correlation function $\xi_{\eta}(\Delta t)$ is related to the power spectral density (PSD) $P_{\eta}(\omega)$ by a 1D Fourier transform
\begin{equation}
\xi_{\eta}(\Delta t) = \int_{-\infty}^{\infty} \dfrac{d\omega}{2\pi} e^{-i \omega \Delta t} P_{\eta}(\omega).
\end{equation}
Since a gaussian random field is fully specified by its PSD, we can forward model $\eta(t)$ with a prescribed functional form of $P_{\eta}(\omega)$. Note that this treatment of the bursty component also assumes that $\xi_{\eta}(\Delta t)$ or $P(\omega)$ is time-invariant, which is a reasonable approximation given the young cosmic age at high redshift for galaxies to evolve significantly in terms of their star formation mode. Following \cite{CT2019} and \cite{PK2023}, we model the PSD as a broken power law that results from a damped random walk \cite{Kelly2014}, 
\begin{equation}
P_{\eta}(\omega) = \frac{\sigma^2_\mathrm{int}}{1/\tau^2_\mathrm{decor}+\omega^2} = \frac{\sigma^2}{1 + (\tau_\mathrm{decor} \omega)^2},
\label{eq:psd_ori}
\end{equation}
where the amplitude (and the long-term variability as $\omega \rightarrow 0$) of fluctuations is set by $\sigma = \sigma_\mathrm{int} \tau_\mathrm{decor}$ and $\tau_\mathrm{decor}$ characterizes the timescale over which fluctuations of $\eta$ decorrelate. To capture the physics that can lead to different `strengths' of decorrelation, we adopt the generalized form of eq. (\ref{eq:psd_ori}) with the frequency dependence being set by an additional parameter $\alpha$ that determines the power-law slope of the PSD, 
\begin{equation}
P_{\eta}(\omega) = \frac{\sigma^2}{1 + (\tau_\mathrm{decor} \omega)^{\alpha}}.
\label{eq:psd}
\end{equation}
Eq.~(\ref{eq:psd}) specifies the 3-parameter PSD model that we use in this work to study bursty star formation and its constraints from galaxy summary statistics. 

Following these steps for parameterizing the SFH, we generate SFHs with varying levels of burstiness using the broken power-law PSD model. When choosing the model parameters, we specifically aim to match the UV variability in our two models, while allowing the H$\alpha$ variability to differ significantly accordingly with the PSD. This is achieved by simultaneously adjusting both the shape and amplitude of the PSD in order to obtain SFHs with similar long-term but different short-term variations. 

\begin{table}[tbp]
\centering
\begin{tabular}{|c|c|c||c|c|c|}
\hline
Model & Parameter & Value & Model & Parameter & Value \\
\hline
A & $f_{\star,0}$ & 0.015 & B & $f_{\star,0}$ & 0.015 \\
A & $\sigma$ & 1.9 & B & $\sigma$ & 3.3 \\
A & $\alpha$ & 1 & B & $\alpha$ & 3 \\
A & $\tau_\mathrm{decor}$ & 10\,Myr & B & $\tau_\mathrm{decor}$ & 30\,Myr \\
\hline
\end{tabular}

\caption{\label{tab:i}SFE and PSD parameters assumed for our two references models.}
\end{table}

In figure~\ref{fig:sfh-psd}, we explicitly show that a strongly time-variable SFH can be obtained by supplementing a smooth SFH with a gaussian random variable that describes burstiness as in eq.~(\ref{eq:sfh}). The characteristics of SFR fluctuations, including both their amplitude and temporal correlation, are specified by the normalization and shape of the PSD. Specifically, we consider two models, Model~A and Model~B, for our subsequent analysis, for which we assume $\{ f_{\star,0}, \sigma, \alpha, \tau_\mathrm{decor} \} = \{0.015, 1.9, 1, 10\,\mathrm{Myr}\}$ and $\{0.015, 3.3, 3, 30\,\mathrm{Myr}\}$, respectively, as summarized in table~\ref{tab:i}. From the comparison of the SFHs and PSDs shown in the top and bottom panels, it is clear that varying the shape of the PSD set by $\tau_\mathrm{decor}$ and $\alpha$ directly impacts the way SFR fluctuations correlate in time. On the other hand, varying the PSD normalization by $\sigma$ sets the amplitude of SFR fluctuations. Increasing the power-law slope $\alpha$ and/or the decorrelation timescale $\tau_\mathrm{decor}$ has the effect of placing more weight on long-term rather than short-term SFR fluctuations, yielding a SFH that is smoother on short timescales.

\subsection{Galaxy SED}

With the SFH in hand, we can take the corresponding stellar age distribution and forward model the galaxy SED via stellar population synthesis (SPS), which creates a synthetic galaxy spectrum from spectral templates of individual simple stellar populations (SSPs) of different age, $t_\mathrm{age}$, and metallicity, $Z$. Specifically, the SPS procedure that yields the specific luminosity at a given wavelength $\lambda$ can be expressed as \cite{Johnson2021}
\begin{equation}
L_{\lambda} = \int_{0}^{t_\mathrm{age}} \mathrm d t \int_{Z_\mathrm{min}}^{Z_\mathrm{max}} \mathrm{SFR}(t, Z) s_{\lambda}(t, Z) e^{-\tau_{\lambda}(t, Z)} \phi(Z) \mathrm d Z,
\end{equation}
where $s_{\lambda}$ is the spectral template, $\tau_{\lambda}$ is the optical depth of dust attenuation, and $\phi(Z)$ is the probability distribution of $Z$. Note that one can get rid of the metallicity integral by assuming a fixed metallicity throughout. Due to the complicated manner in which bursty star formation may impact the metallicity, we defer a more thorough analysis to future work and simply assume a constant metallicity $Z = 0.1\,Z_{\odot}$ in this paper. Dust attenuation can also be complicated by burst cycles of star formation through e.g., dusty outflows driven by stellar feedback \cite{Kannan2021,Ferrara2024}. For simplicity, we ignore the connection between burstiness and dust here and focus our subsequent analysis on solely the effects caused by bursty star formation. We should emphasize that this is a deliberate simplification given the scope of this work---in practice, dust attenuation, among several other factors, may introduce additional stochasticity in the galaxy SED and modulate UV and H$\alpha$ emission in ways beyond what can be accurately described by simple, semi-empirical methods \cite{Sabti2022,Mirocha2023}. These potential sources of confusion will be discussed in more detail in section~\ref{sec:discussion}, though we leave a quantitative investigation of their impact to future studies. For the SSP templates, we take data products of the Binary Population and Spectral Synthesis (BPASS) v1.0 code \cite{ES2009BPASS} that have been processed by the photoionization and radiative transfer code Cloudy \cite{Ferland2017} to account for the nebular emission (including both lines and continuum). A Chabrier stellar initial mass function \cite{Chabrier2003} and single star spectra are adopted. For each galaxy, individual realizations of its SFH given by our PSD-based method are supplied to the SPS procedure to draw samples of the observables of interest ($L_\mathrm{H\alpha}$ and $L_\mathrm{UV}$). For any given PSD model, we then approximate the resulting probability distribution of the observable, $\mathcal{P}(L_{\lambda})$, with the kernel density estimation (KDE) method assuming gaussian kernels.

\subsection{Summary statistics of the galaxy population}

Given $\mathcal{P}(L_{\lambda})$, we can calculate the observable of interest, such as the galaxy summary statistics in our case, by convolving it with $\mathcal{P}(L_{\lambda})$ (see also \cite{CT2019}). Note that, unlike in many previous studies, here we do not assume a log-normal distribution for $\mathcal{P}(L_{\lambda})$, or a gaussian in $\log L_{\lambda}$, specified by a mean value and a scatter (e.g., $\sigma_{\rm UV}$). Rather, we consider the actual distribution sampled from the PSD realizations and estimated by the KDE. As shown in appendix~\ref{sec:appendix:a}, this avoids the systematic error on the summary statistics associated with the assumption of a gaussian distribution. 

\subsubsection{Luminosity function}
Following \cite{Munoz2023}, we can express the LF as
\begin{equation}
\phi_{\mathcal{O}} = \frac{\mathrm d n}{\mathrm d \mathcal{O}} = \int \mathrm d M_\mathrm{h} \frac{\mathrm d n}{\mathrm d M_\mathrm{h}} \mathcal{P}(\mathcal{O} | M_\mathrm{h}), 
\label{eq:lf}
\end{equation}
where $\mathrm d n / \mathrm d M_\mathrm{h}$ is the halo mass function \cite{ST1999} and the observable $\mathcal{O}$ is taken to be either the 1500\AA\ UV continuum magnitude, $M_\mathrm{UV}$, or the H$\alpha$ luminosity, $L_\mathrm{H\alpha}$, in this paper. 

\subsubsection{Clustering}

Following \cite{Munoz2023}, we consider the (number-weighted) bias of galaxies selected by the observable of interest as the key metric for clustering, which can be expressed as
\begin{equation}
b_{\mathrm{eff},\mathcal{O}} = \phi^{-1}_\mathcal{O} \int \mathrm d M_\mathrm{h} \frac{\mathrm d n}{\mathrm d M_\mathrm{h}} b(M_\mathrm{h}) \mathcal{P}(\mathcal{O} | M_\mathrm{h}), 
\label{eq:b}
\end{equation}
where $b(M_\mathrm{h})$ is the halo bias factor. 

\subsection{A Fisher-matrix framework for SFH model constraints} \label{sec:models:fisher}

Considering the posterior distribution, $P(\theta|\mathcal{S})$, of SFH model parameters $\boldsymbol{\theta}$ given the observed galaxy summary statistics $\mathcal{S}$ (where $P$ is assumed to be gaussian and the $\mathcal{S}$ uncorrelated), we can derive the covariance matrix $C(\boldsymbol{\theta})$, using the Fisher matrix,
\begin{equation}
F_{ij} = - \left\langle \frac{\partial^2 \ln P}{\partial \theta_i \partial \theta_j} \right\rangle = \sum_{\mathcal{O}} \frac{1}{\mathrm{var}[\mathcal{S(O)}]} \frac{\partial \mathcal{S(O)}}{\partial \theta_i} \frac{\partial \mathcal{S(O)}}{\partial \theta_j},
\end{equation}
which gives
\begin{equation}
F^{-1} = C =  
  \left[ {\begin{array}{ccc}
    \mathrm{var}(\theta_0) & \cdots & \mathrm{cov}(\theta_0,\theta_N) \\
    \vdots & \ddots & \vdots \\
    \mathrm{cov}(\theta_0,\theta_N) & \cdots & \mathrm{var}(\theta_N) \\
  \end{array} } \right]. 
\end{equation}
In what follows, we consider a 4-parameter model of the SFH specified by $\boldsymbol{\theta} = \{ f_{\star,0}, \sigma, \tau_\mathrm{decor}, \alpha \}$. The first parameter, $f_{\star,0}$, sets the overall normalization of the SFE as defined by the double power-law model in eq. (\ref{eq:sfe}), whereas the other 3 parameters define the PSD. For simplicity, we hold the rest of SFE parameters ($M_\mathrm{p}$, $\gamma_\mathrm{lo}$, and $\gamma_\mathrm{hi}$) fixed at their best-fit values from \cite{Mirocha2017} and ignore the redshift evolution of all parameters. It is also noteworthy that the mass dependence of the PSD is likely non-negligible in reality due to e.g., the transition from bursty to steady star formation as galaxies grow massive enough and become more stable against feedback \cite{FM2022,Hopkins2023}. Here, we choose to stick with the simple 4-parameter model in this proof-of-the-concept study and postpone a more flexible parameterization to future work. 

To create mock data for a given model and survey specification (as input for the Fisher matrix analysis), we follow the method in section 4 of \cite{Munoz2023} to derive the signal-to-noise ratios of the summary statistics of interest. Specifically, for the LF, we combine in quadrature the error contributions from Poisson noise ($\propto \sqrt{N_\mathrm{gal}}$, where $N_\mathrm{gal}$ is the number of galaxies in each luminosity bin) and cosmic variance, with the latter being estimated by the analytic method introduced in \cite{TF2020}. For the effective bias, we calculate its uncertainty from the detectability of the clustering angular power spectrum using the Knox formula \cite{Knox1995}. 

\section{Results} \label{sec:results}

\subsection{Impact of the burstiness on SFR indicators}

\begin{figure}[htbp]
\centering
\includegraphics[width=0.825\textwidth]{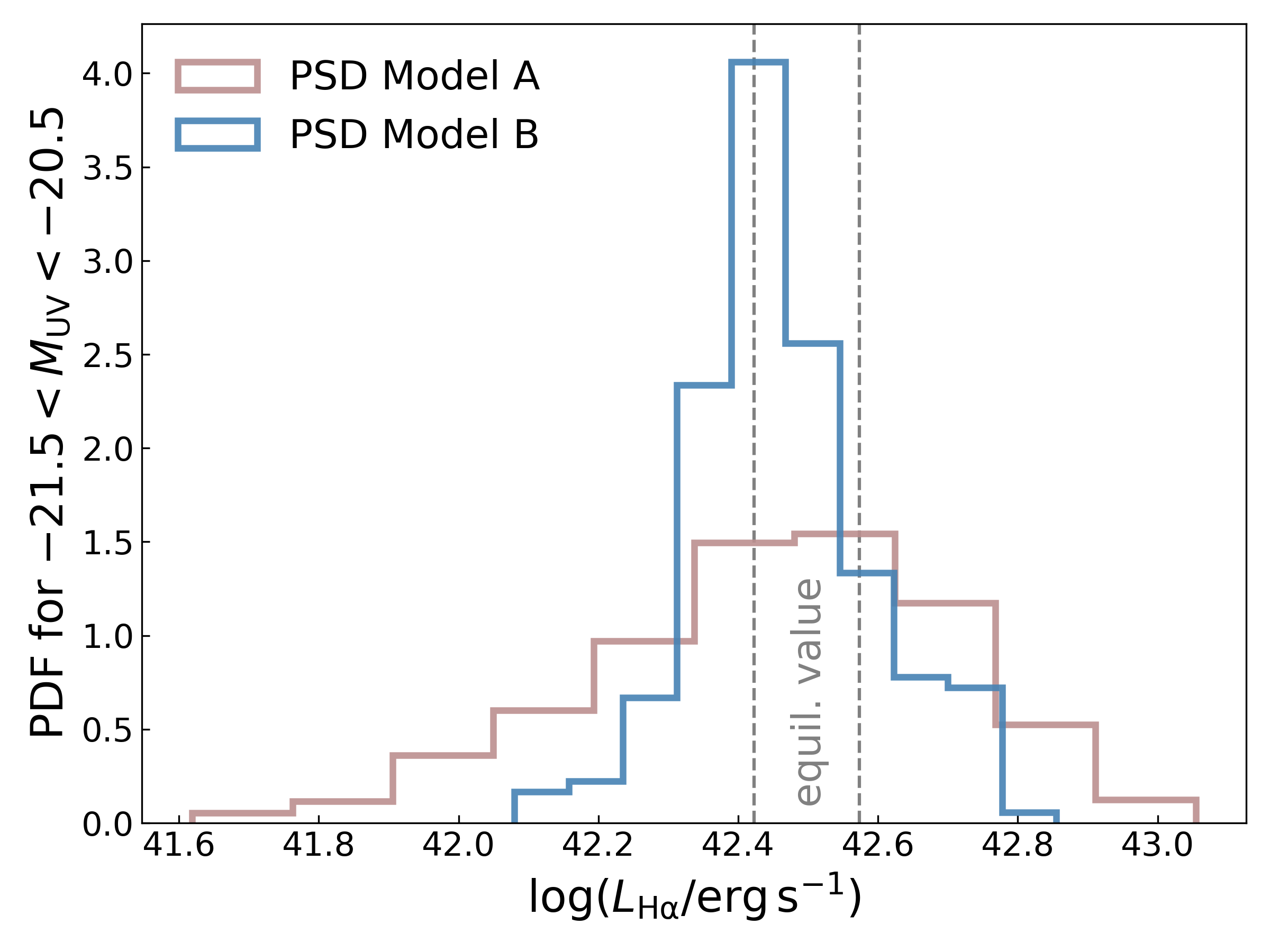}
\caption{The distribution of H$\alpha$ luminosity for a given narrow range of UV magnitude $-21.5 < M_\mathrm{UV} < -20.5$ in our two PSD models. The distribution significantly widens when the PSD predicts SFHs with a higher level of burstiness on short timescales, as in Model~A. For comparison, the equilibrium value corresponding to $M_\mathrm{UV}=-21$ from \cite{Asada2024} for a H$\alpha$-to-UV luminosity ratio between 1/85 and 1/60 is plotted.}
\label{fig:ha_vs_uv}
\end{figure}

While UV continuum and H$\alpha$ are both commonly used SFR indicators, they trace different star formation timescales (a few Myr for H$\alpha$ and 10--100\,Myr for UV) due to the different ways they are produced \cite{JFV2021}. As a result, changes in the level of burstiness prescribed by the PSD can lead to distinct joint distributions of these SFR indicators. To illustrate this effect, we show in figure~\ref{fig:ha_vs_uv} the distribution of H$\alpha$ luminosity for a narrow range of UV magnitude $-21.5 < M_\mathrm{UV} < -20.5$ in our two PSD models A and B. For comparison, we also plot the equilibrium value corresponding to $M_\mathrm{UV} = -21$ converted from a H$\alpha$-to-UV luminosity ratio between 1/85 and 1/60, which is derived assuming a smooth and steady SFH with plausible variations of the stellar metallicity and IMF \cite{Mehta2023,Asada2024}. The widened distribution of $L_\mathrm{H\alpha}$ in Model~A compared to Model~B indicates that the contrast between these SFR indicators is a useful probe for bursty SFHs---a method that has been extensively explored in the literature to quantify the burstiness for different galaxy populations using either galaxy-integrated or spatially resolved observations \cite{Emami2019,Faisst2019,JFV2021,Sun2023,Asada2024,Clarke2024}. This motivates our investigation of how to constrain bursty star formation using galaxy summary statistics, such as the LF and clustering bias, measured in these SFR indicators. 

\begin{figure}[htbp]
\centering
\includegraphics[width=\textwidth]{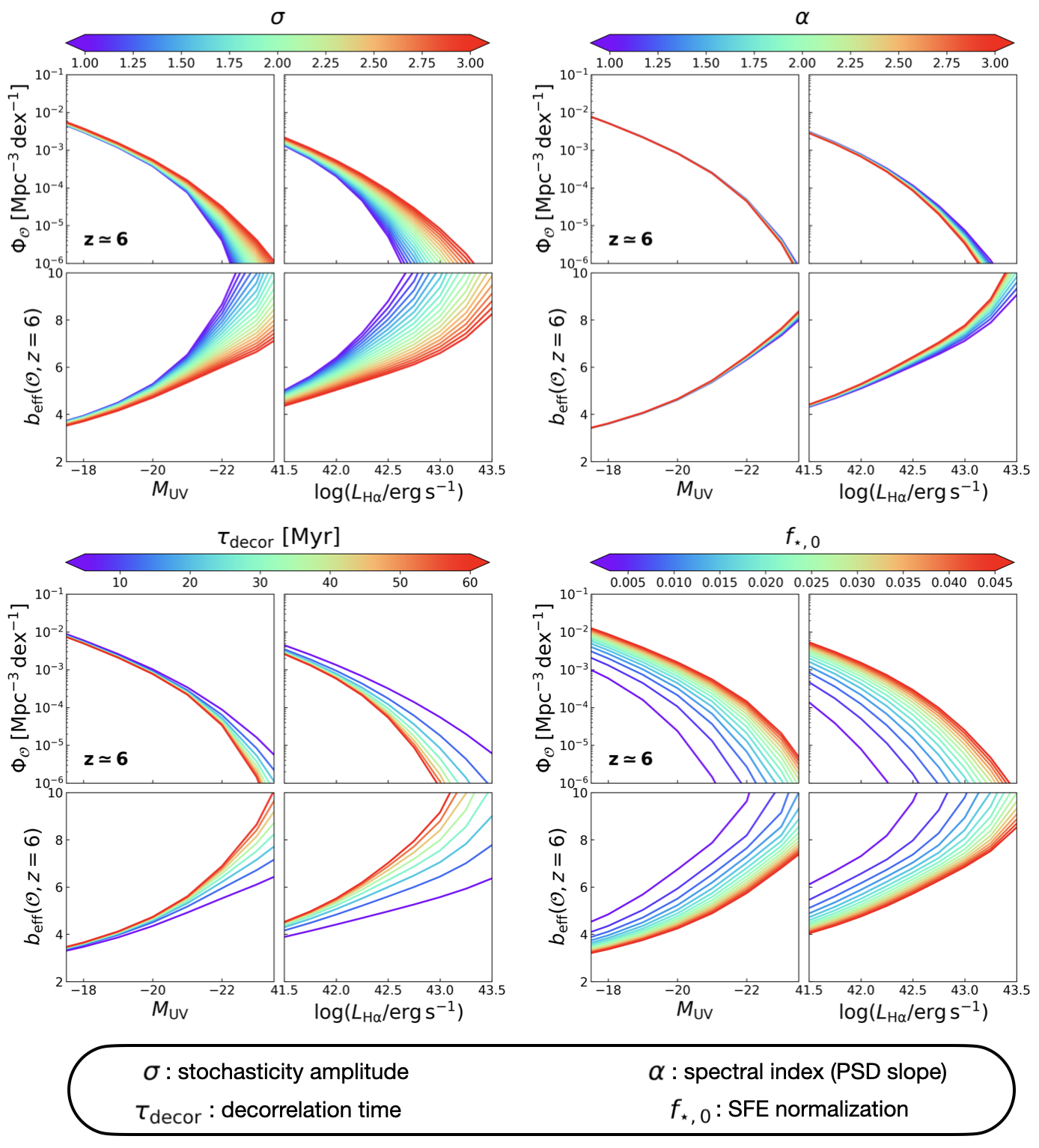}
\caption{Visualization of the effects of changing the three PSD model parameters and one SFE parameter on the galaxy summary statistics. The correlation patterns of these parameters can also been seen from the color coding. In each panel, only one of the four free parameters is varied as shown by the color coding while the other three are held fixed. The higher sensitivity of H$\alpha$ to the temporal correlation of SFR fluctuations is clearly visible through the relevant parameters $\alpha$ and $\tau$, especially for short timescales 5--10\,Myr typically traced by H$\alpha$ emission.}
\label{fig:visualize}
\end{figure}

\subsection{Combining SFR indicators tracing different timescales}

\subsubsection{Responses of summary statistics to PSD parameters}

To understand the exact impact of bursty SFHs on galaxy summary statistics and the prospects for using them to constrain burstiness, we follow the steps in section~\ref{sec:models} to derive the one-point (LF) and two-point (clustering bias) statistics from the luminosity/magnitude distributions implied by the SFH realizations drawn from the PSD model. For each combination of PSD parameters, we generate 1000 random realizations of the SFH along with the corresponding galaxy SED given by the SPS. We reiterate that because the full distribution $\mathcal{P}(\mathcal{O})$, where the observable $\mathcal{O}$ is either $M_\mathrm{UV}$ or $L_\mathrm{H\alpha}$, is directly sampled by our random realizations of the SFH and galaxy SED, we do not need to assume an ad hoc functional form (which is usually taken to be gaussian in the literature) for $\mathcal{P}(\mathcal{O})$ to perform the convolutions in eq. (\ref{eq:lf}) and eq. (\ref{eq:b}).  

It is instructive to first build intuition of how each PSD model parameter or the SFE normalization parameter impacts the galaxy summary statistics by varying them in a controlled manner. In figure~\ref{fig:visualize}, we visualize the response of the UV and H$\alpha$ LFs and bias factors to each parameter by continuously varying one of the four parameters at a time while keeping the other three fixed. These results clearly demonstrate that the temporal correlation of SFR fluctuations characterized by parameters $\alpha$ and $\tau$ can be better probed by the H$\alpha$ statistics, especially by the bright emitters and for timescales H$\alpha$ emission is most sensitive to ($\lesssim 10\,$Myr). Additionally, the parameter dependence of the summary statistics shown by the color-coding indicates the degeneracy direction of these parameters. 

\begin{figure}
\centering
\includegraphics[width=\textwidth]{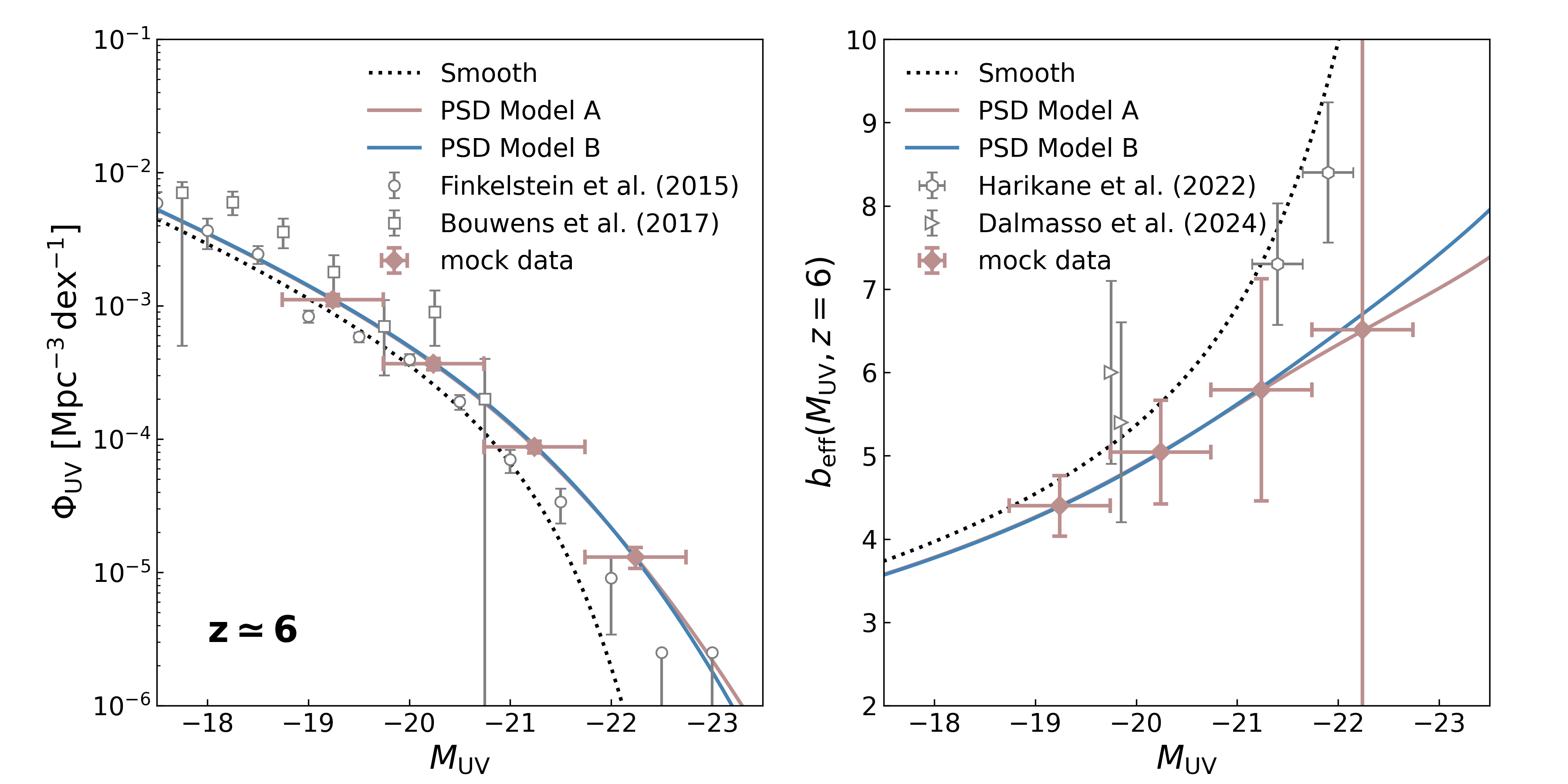}
\caption{The galaxy LF and number-weighted clustering bias evaluated for $M_\mathrm{UV}$ at $z=6$. Predictions from PSD model variations implying different levels of burstiness in the SFH (solid curves) are contrasted against that based on a non-bursty, smooth SFH (dotted curve). Since the Models~A and B predict comparable levels of UV variability, they remain largely indistinguishable by UV summary statistics. Mock data used for Fisher matrix analysis are generated assuming a hypothetical JWST/NIRCam survey covering 0.6\,deg$^{2}$ with $m_\mathrm{AB}^\mathrm{lim}=28$ (displayed only for Model~A). For comparison, existing observational constraints on the UV LF and clustering bias at $z\simeq6$ are plotted \cite{Finkelstein2015,Bouwens2017,Harikane2022,Dalmasso2024}.}
\label{fig:stats_uv}
\end{figure}

\begin{figure}
\centering
\includegraphics[width=\textwidth]{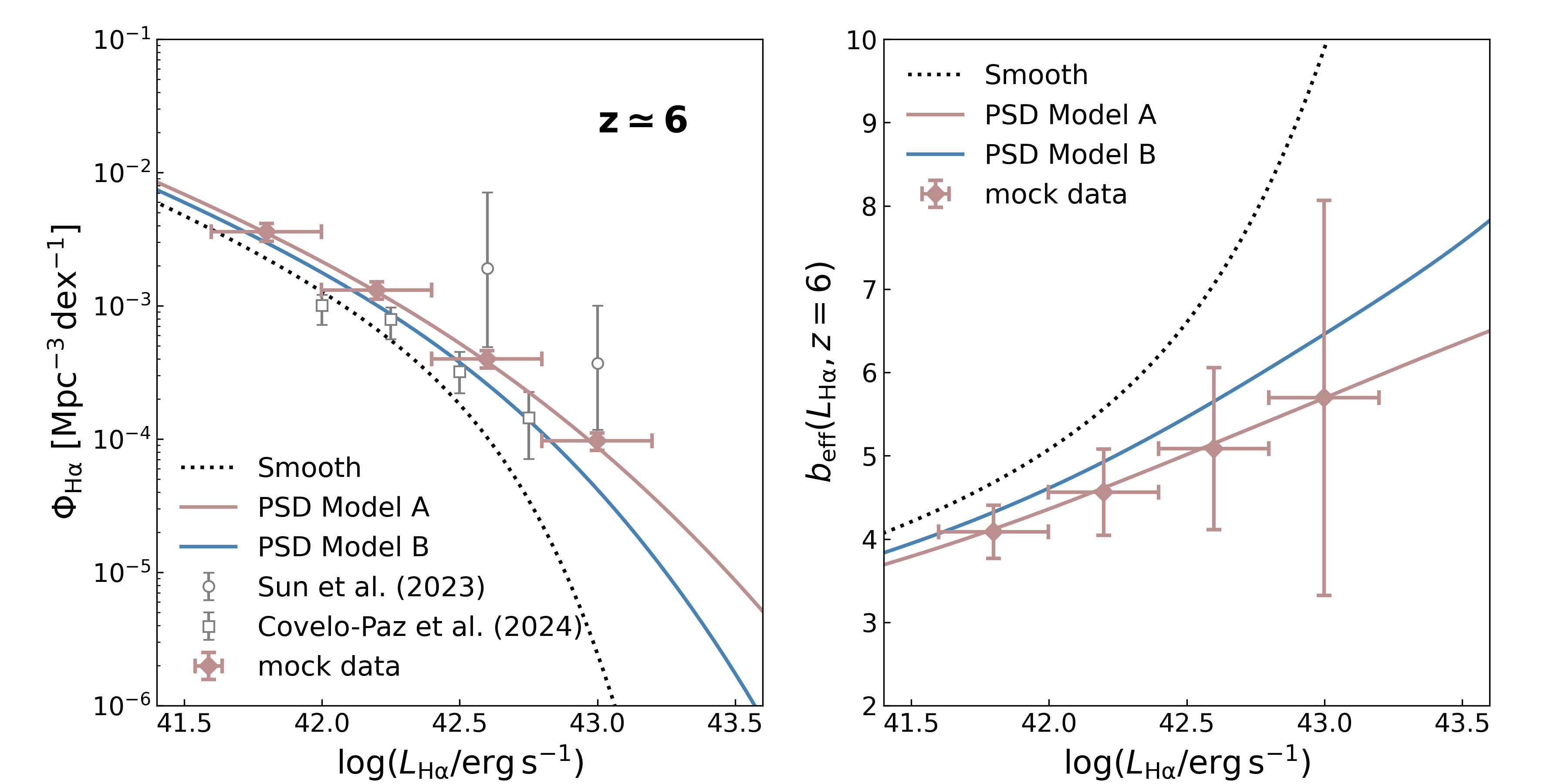}
\caption{Same as figure~\ref{fig:stats_uv} but for H$\alpha$ luminosity. Note that the two models were largely indistinguishable in UV summary statistics but depart significantly for $L_\mathrm{H\alpha}$-based statistics thanks to its sensitivity to SFR variations on shorter timescales. Mock data used for Fisher matrix analysis are generated assuming a hypothetical JWST/NIRCam survey covering 0.6\,deg$^{2}$ with $m_\mathrm{AB}^\mathrm{lim}=28$ (displayed only for Model~A). For comparison, existing observational constraints from JWST on the H$\alpha$ LF at $z\simeq6$ are plotted \cite{Sun2023HALF,Covelo-Paz2024}.}
\label{fig:stats_ha}
\end{figure}

\subsubsection{Comparison of model variations and observations}

To further elaborate on the complementarity of H$\alpha$ to UV continuum emission in this context, in figure~\ref{fig:stats_uv}, we show at $z = 6$ the UV LF and the number-weighted clustering bias factor as a function of $M_\mathrm{UV}$ obtained from the two different PSD models, along with a model that contains only the smooth SFH component (i.e., no burstiness). As a sanity check, we also compare our model predictions to some observational constraints on these summary statistics. It is evident that bursty SFHs lead to a flattened LF and a reduced clustering bias at the bright end ($M_\mathrm{UV} \lesssim -21$), both of which are consequences of the up-scattering of more abundant low-mass halos to higher luminosities. On the other hand, because all of the 2 PSD model variations are designed to yield comparable UV variability, their corresponding summary statistics appear very similar and hardly distinguishable despite the vastly different SFHs in these models (see figure~\ref{fig:sfh-psd}). 

The degeneracy of these scenarios of bursty star formation for galaxy summary statistics based on $M_\mathrm{UV}$ calls for another probe of the SFR variability on a different timescale. In figure~\ref{fig:stats_ha}, we show that considering similar summary statistics but for $L_\mathrm{H\alpha}$ can make it significantly easier to distinguish these SFH models. The left panel of figure~\ref{fig:stats_ha} shows a comparison of the $\mathrm{H\alpha}$ LF, $\Phi_\mathrm{H\alpha}$, predicted by the PSD models with a reference model without burstiness, as well as recent observational constraints on $\Phi_\mathrm{H\alpha}$ at $z\sim6$ from JWST. Notably, the inclusion of burstiness in the SFH provides the enhancement of the bright-end $\Phi_\mathrm{H\alpha}$ required to match the observations. The predicted clustering bias as a function of $L_\mathrm{H\alpha}$ is shown in the right panel of figure~\ref{fig:stats_ha}. Unlike the case of $M_\mathrm{UV}$, here the bias factors in different PSD model variations differ much more significantly due to the different $\mathrm{H\alpha}$ variability predicted by them as figure~\ref{fig:ha_vs_uv} indicates. 

Together, figures~\ref{fig:stats_uv} and \ref{fig:stats_ha} allude to the power of combining the one-point and two-point statistics of galaxies measured in both $M_\mathrm{UV}$ and $L_\mathrm{H\alpha}$. Not only does it reduce the degeneracy between the burstiness-induced UV variability and an enhanced star formation efficiency as proposed in \cite{Munoz2023}, but it also allows detailed information about the bursty SFH to be studied, such as the amplitude and temporal correlation of SFR fluctuations encoded in our Models A and B. 

\begin{figure}[htbp]
\centering
\includegraphics[width=\textwidth]{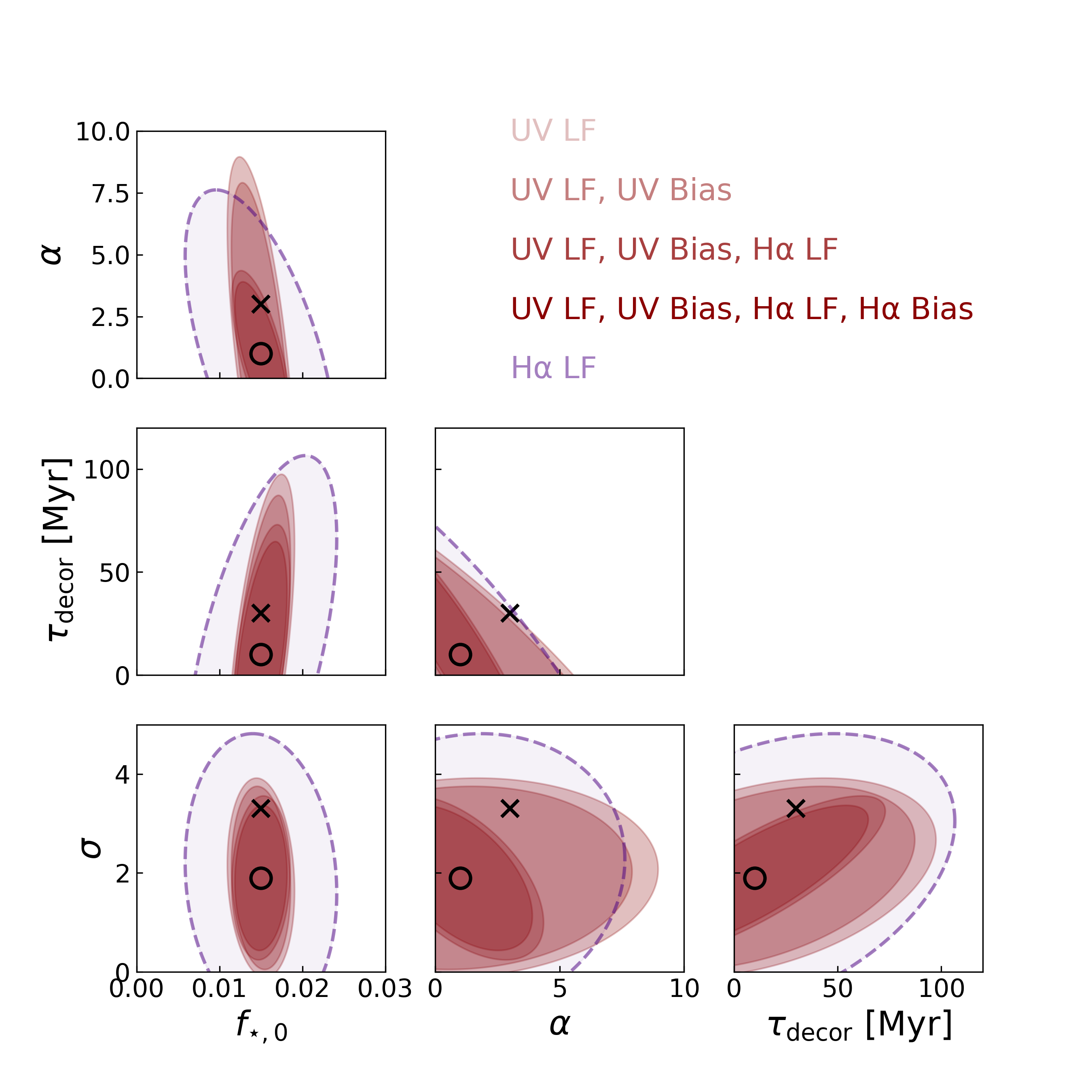}
\caption{1-sigma constraints on the SFE and PSD parameters in Model A derived from the mock LF and clustering observations shown in figures~\ref{fig:stats_uv} and \ref{fig:stats_ha} using the Fisher matrix, assuming a gaussian prior on $\alpha$ with $\Delta \alpha = 5$ (1-sigma). In total, 5 ellipses are shown in each panel, which correspond to the cases when using only the UV LF, only the H$\alpha$ LF, both the UV LF and bias, all but the H$\alpha$ bias, and all 4 summary statistics, respectively. To show how well the two reference models can be distinguished, truth values of the parameters in Models~A and B are marked as circles and x's, respectively. Note that the unphysical portion of the ellipses is not shown given that all the 4 model parameters are non-negative. }
\label{fig:fm}
\end{figure}

\subsubsection{Fisher matrix forecasts}

Also overplotted in figures~\ref{fig:stats_uv} and \ref{fig:stats_ha} are the error bars on the LF and clustering bias factor per magnitude/luminosity bin, estimated for a hypothetical JWST/NIRCam multi-band survey covering an area of 0.6\,deg$^2$ with a limiting magnitude for source detection $m^\mathrm{lim}_\mathrm{AB}=28$ (inspired by COSMOS-Web~\cite{Casey2023}). The H$\alpha$ luminosity is assumed to be measured from the medium-band photometry, similar to the methodology used in \cite{Asada2024}. To account for observational uncertainties and systematics not captured by our estimates of the Poisson noise and cosmic variance (see section~\ref{sec:models:fisher}), we conservatively impose a minimum fractional error of 10\% and 20\% per bin for continuum UV and H$\alpha$ measurements, respectively. The resulting constraints on the summary statistics are used to evaluate the Fisher matrix. 

We show in figure~\ref{fig:fm} the joint 1-sigma constraints on the SFE and PSD model parameters predicted by the Fisher matrix, assuming mock observations of the galaxy summary statistics for Model A as shown in figures~\ref{fig:stats_uv} and \ref{fig:stats_ha}. We note that, as shown in figure~\ref{fig:visualize}, even H$\alpha$ statistics are only mildly sensitive to changes in $\alpha$. However, since further increasing $\alpha$ after the PSD becomes sufficiently steep would essentially result in the same sharp cutoff at $\tau_\mathrm{decor}$, to avoid $\alpha$ to become exceedingly large we impose a gaussian prior on $\alpha$ with a standard deviation of $\Delta \alpha = 5$, motivated by estimates of $\alpha$ by \cite{Iyer2020}. In total, five sets of ellipses corresponding to different combinations of summary statistics, along with truth values of the parameters in both Models A (circles) and B (x's), are plotted to illustrate how the addition of new statistics reduces parameter degeneracies, tightens constraints, and better distinguishes models. 

As is evident from the improvement on the parameter constraints after including the H$\alpha$ LF (and bias) measurements, the UV data is not constraining enough for the $\alpha$ and $\tau_\mathrm{decor}$ parameters quantifying the temporal correlation of SFR fluctuations. This is more clearly shown in the 3 bottom panels focusing on the joint distribution of PSD parameters themselves. Measurements of the H$\alpha$ summary statistics are critical for achieving meaningful constraints on the SFR variability on different timescales. When all 4 summary statistics are included in the analysis, for Model~A we get the following 1-sigma constraints on $f_{\star,0} \approx (15 \pm 2.2) \times 10^{-3}$ and $\sigma \approx 1.9 \pm 1.0$ (1-sigma), together with and 3-sigma upper limits $\alpha \lesssim 6$ and $\tau_\mathrm{decor} \lesssim 100\,$Myr (also note that all these parameters must be non-negative to be physical). With the inclusion of H$\alpha$ statistics, the resulting joint parameter distributions (especially that between $\tau_\mathrm{decor}$ and $\alpha$) make it possible to distinguish Models~A and B, which are otherwise similarly favored by the UV data. We also note that when only the LF constraints are used, figure~\ref{fig:fm} suggests significantly less degeneracy between $\sigma$ and $f_{\star,0}$ than what \cite{Munoz2023} found. This is mainly because the effects of these two parameters on the LF are not completely degenerate (see figure~\ref{fig:visualize}), which leave them distinguishable especially when the bright-end LF can be well measured. In practice, when uncertainties in other aspects of the SFE, such as the mass dependence captured by $M_\mathrm{p}$, $\gamma_\mathrm{lo}$, and $\gamma_\mathrm{hi}$ (which we hold fixed), are taken into account, the effects of burstiness and the SFE on the LF would become more degenerate. 

\section{Discussion}
\label{sec:discussion}

So far, our proof-of-the-concept analysis has demonstrated the great potential of combining the galaxy LF and clustering measured for different SFR indicators to probe bursty SFHs of high-$z$ galaxies. However, there are a few noteworthy caveats and limitations that motivate further extensions of the presented framework in the future. 

The usage of galaxy summary statistics is complicated by uncertainties associated with the SFR indicators themselves. The impact of dust attenuation is arguably the most important complicating factor, given the challenge of accurately determining the amount of dust correction required for the observed UV and H$\alpha$ luminosities. Uncertainties arising from measurements of dust reddening, the dust attenuation law assumed (which can vary across the galaxy population), and differential dust attenuation due to distinct dust distributions with respect to differently aged stellar populations can all result in modulations of the observed UV and H$\alpha$ statistics not associated with bursty star formation \cite{Shivaei2015,Shapley2023}. In this work, we avoid the complications of dust entirely by working only with the intrinsic UV and H$\alpha$ emission. This is mainly because our simplistic model cannot adequately predict how dust may complicate the summary statistics beyond what can be expected from simple estimates based on empirical scaling relations. In reality, there can be significant galaxy-to-galaxy and/or time variations of dust attenuation for UV and H$\alpha$, which can be caused by e.g., different dust distributions and burst cycles of star formation may clear up interstellar dust by driving dusty outflows via stellar feedback \cite{Ferrara2024,Ferrara2024z14}. More sophisticated physical models, including ones that build on insights from numerical simulations with realistic galaxy models and detailed radiative transfer calculations \cite{Katz2019,Kado-Fong2020,Tacchella2022}, are thus necessary for reliably determining the amount of dust correction needed for galaxies with strongly bursty SFHs. This is particularly true considering that dust effects can in principle be correlated with a galaxy's burst cycles. Consequently, we stress that the constraining power on bursty SFHs from the summary statistics shown in this work should be taken as optimistic estimates when dust effects can be fully corrected for. 

A number of other factors related to the stellar population such as the IMF, stellar metallicity, binary fraction, and their variations across the galaxy population can also lead to complications by modifying the intrinsic production of UV and H$\alpha$ emission and thereby changing their scatters \cite{Katz2019,Wilkins2020}. Since analyses have shown that the joint distribution of dust-corrected UV and H$\alpha$ luminosities for galaxies at lower redshift may not be explained by the stochastic sampling and/or time evolution of these factors \cite{Eldridge2012,Emami2019}, they are likely less of a concern than dust for constraining bursty SFHs with UV and H$\alpha$ emission. Additional factors such as contributions from non-stellar sources (e.g., AGN and shocks) and the environmental dependence can also make the connection between UV/H$\alpha$ emission and the SFR more sophisticated \cite{Peters2017}, although they can be partially circumvented by careful sample selection. Nevertheless, more observations will be needed to better understand and test the theoretical predictions of all these effects, especially at high redshift. In future studies, it would be interesting to develop empirically based extensions of the current dust-free framework to characterize how dust and properties of the stellar population impact the observed UV and H$\alpha$ luminosities and the corresponding galaxy summary statistics. 

Though simple, our framework is highly flexible in terms of both the form of bursty SFHs, which can be modeled with different parameterization of the PSD, and the observables to predict, which are not limited to the one-point and two-point statistics of UV continuum and H$\alpha$ emission. Such flexibility allows it to be readily extended to investigate e.g., the potential dependence of burstiness on galaxy mass \cite{Weisz2012,KB2024} and more sophisticated functional forms of the PSD, where multiple characteristic timescales are involved and associated with different physical processes driving the burst cycles (e.g., stellar feedback and gas recycling) \cite{Iyer2024}. Additional diagnostics of the SFH, including other emission lines like [OIII], the UV continuum slope, and the Balmer break, can be easily incorporated into the analysis as well. It is also relatively straightforward to apply our semi-analytic framework to forward model metrics of burstiness for individual systems binned by galaxy properties (e.g., stellar mass), thereby including such observations in the parameter
inference. In future works, it would also be valuable to utilize the simplicity and flexibility of this framework to further investigate the connection between bursty SFHs and the galaxy summary statistics. Given the importance and wide usage of these galaxy summary statistics for simultaneously probing both galaxy astrophysics and cosmological models \cite{Schive2016,Sabti2022,Dayal2024,Sipple2024}, it is crucial to better understand how this connection enables (and requires) the astrophysics of star formation and the underlying cosmology to be jointly constrained. 

\section{Conclusions} \label{sec:conclusions}

We have presented a flexible semi-analytic modeling framework for constraining bursty SFHs from the galaxy summary statistics, specifically the galaxy LF and clustering measured for UV continuum and H$\alpha$ emission. By describing the bursty component of the SFH as a gaussian random variable that can be specified by a parametric model of the PSD, we employ this framework to self-consistently compute galaxy SEDs and sample the probability distribution of the observables of interest ($M_\mathrm{UV}$ and $L_\mathrm{H\alpha}$). Based on the resulting distributions, we can derive the galaxy LF and clustering signals of interest in different scenarios of bursty star formation specified by the PSD. 

As a proof of concept, we have combined this framework with the Fisher matrix method to forecast the parameter constraints for a generic, 4-parameter model of bursty SFHs for galaxies at $z\sim6$. By considering UV and H$\alpha$ LFs and clustering bias factors measurable from a hypothetical JWST/NIRCam survey similar to COSMOS-Web, we have shown that combining the 1-point and 2-point statistics of UV and H$\alpha$ allows PSD model parameters to be separated from the SFE and individually constrained as either detections or upper limits. Thanks to the short star formation timescales it probes, H$\alpha$ is crucial for constraining parameters that affect the temporal correlation of bursty SFHs (i.e., $\tau_\mathrm{decor}$ and $\alpha$) when combined with UV. The resulting joint parameter constraints also allow different PSD models to be distinguished from each other, as demonstrated for our reference Models~A and B. 

The framework presented demonstrates the prospects of using the galaxy summary statistics measured for different SFR indicators to probe bursty SFHs of high-$z$ galaxies, provided that potential confusion due to e.g., dust attenuation and stellar population effects can be properly accounted for. It can be readily extended to take into account (1) more sophisticated prescriptions of the SFR variability, such as ones that consider the mass/redshift dependence or multiple correlation timescales, and (2) statistics of a wider range of SFH diagnostics, such as the UV slope and the Balmer break strength. When combined with physical models that predict the relevant statistics, these extensions will make it possible to extract more detailed physics behind stochastic SFHs of galaxies at different cosmic times.

\acknowledgments

We thank the anonymous referee for insightful and constructive comments that that helped improve this paper, as well as Yoshihisa Asada, Ryan Endsley, Viola Gelli, Kartheik Iyer, Harley Katz, and Andrey Kravtsov for stimulating discussions. GS was supported by a CIERA Postdoctoral Fellowship. GS and JBM would like to acknowledge the Kavli Institute for Theoretical Physics (KITP) where part of this work was done for their hospitality, which is supported in part by grant NSF PHY-2309135. JBM was supported by NSF through grants AST-2307354 and AST-2408637. JM was supported by an appointment to the NASA Postdoctoral Program at the Jet Propulsion Laboratory/California Institute of Technology, administered by Oak Ridge Associated Universities under contract with NASA. Part of this work was done at Jet Propulsion Laboratory, California Institute of Technology, under a contract with the National Aeronautics and Space Administration (80NM0018D0004). CAFG was supported by NSF through grants AST-2108230 and AST-2307327; by NASA through grants 21-ATP21-0036 and 23-ATP23-0008; and by STScI through grant JWST-AR-03252.001-A. 

\appendix
\section{Effects of a gaussian approximation on the summary statistics} \label{sec:appendix:a}

\begin{figure}
\centering
\includegraphics[width=0.95\textwidth]{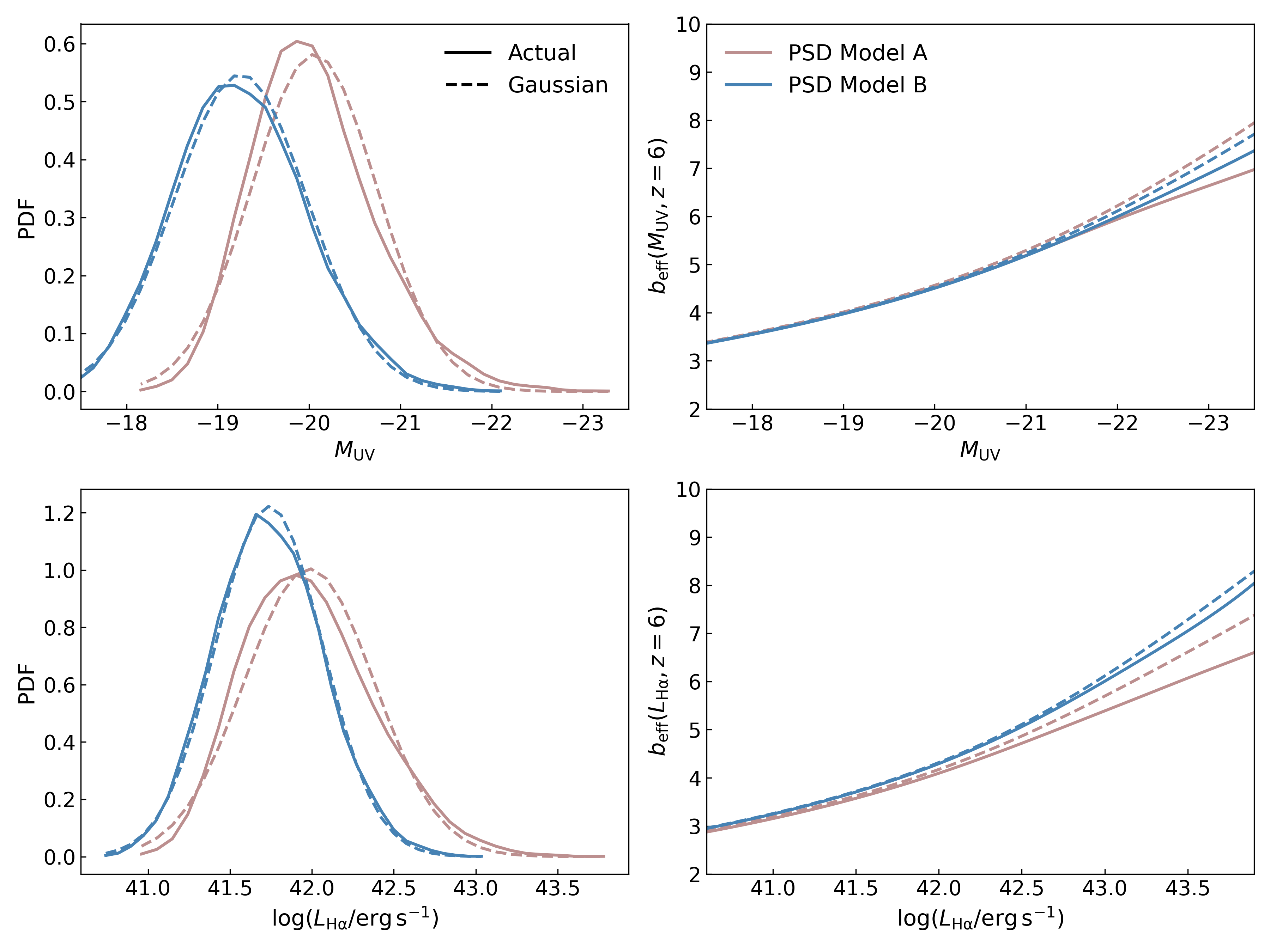}
\caption{The validity of approximating the probability distributions $\mathcal{P}(M_\mathrm{UV})$ and $\mathcal{P}(L_\mathrm{H\alpha})$ with a gaussian. Using the best-fit gaussian tends to overestimate the effective clustering bias, especially for bright galaxies, when the SFH is highly bursty as in Model~A. This is a relatively small effect that can nevertheless be non-negligible for precision measurements and analysis of the summary statistics.}
\label{fig:gaussian_vs_kde}
\end{figure}

In figure~\ref{fig:gaussian_vs_kde}, we assess the validity of 
assuming a gaussian for the probability distributions $\mathcal{P}(M_\mathrm{UV})$ and $\mathcal{P}(\log L_\mathrm{H\alpha})$, which is often done in the literature.
We compare the effective clustering bias $b_\mathrm{eff}$ derived using the actual distribution drawn from the SFH realizations (and approximated by the KDE method) against that derived from a best-fit gaussian. The left panels show that the actual distribution is not a perfect gaussian but slightly skewed to the faint side, especially for Model~A that is more bursty. As a result, $b_\mathrm{eff}$ of bright sources predicted by a gaussian approximation tends to be (up to 10--20\%) higher than what the actual distribution predicts. This offset may become a non-negligible systematic effect when summary statistics like $b_\mathrm{eff}$ can be measured at high precision. 

\bibliographystyle{JHEP}
\bibliography{burst.bib}

\providecommand{\href}[2]{#2}\begingroup\raggedright\begin{thebibliography}{10}

\bibitem{Ciesla2024}
L.~{Ciesla}, D.~{Elbaz}, O.~{Ilbert}, V.~{Buat}, B.~{Magnelli}, D.~{Narayanan} et~al., \emph{{Identification of a transition from stochastic to secular star formation around z = 9 with JWST}}, \href{https://doi.org/10.1051/0004-6361/202348091}{\emph{Astron. Astrophys.} {\bfseries 686} (2024) A128} [\href{https://arxiv.org/abs/2309.15720}{{\ttfamily 2309.15720}}].

\bibitem{Dressler2024}
A.~{Dressler}, M.~{Rieke}, D.~{Eisenstein}, D.P.~{Stark}, C.~{Burns}, R.~{Bhatawdekar} et~al., \emph{{Building the First Galaxies{\textemdash}Chapter 2. Starbursts Dominate the Star Formation Histories of 6 < z < 12 Galaxies}}, \href{https://doi.org/10.3847/1538-4357/ad1923}{\emph{Astrophys. J.} {\bfseries 964} (2024) 150} [\href{https://arxiv.org/abs/2306.02469}{{\ttfamily 2306.02469}}].

\bibitem{Endsley2024}
R.~{Endsley}, D.P.~{Stark}, L.~{Whitler}, M.W.~{Topping}, B.D.~{Johnson}, B.~{Robertson} et~al., \emph{{The star-forming and ionizing properties of dwarf z 6-9 galaxies in JADES: insights on bursty star formation and ionized bubble growth}}, \href{https://doi.org/10.1093/MNRAS/stae1857}{\emph{Mon. Not. Roy. Astron. Soc.} {\bfseries 533} (2024) 1111} [\href{https://arxiv.org/abs/2306.05295}{{\ttfamily 2306.05295}}].

\bibitem{Asada2024}
Y.~{Asada}, M.~{Sawicki}, R.~{Abraham}, M.~{Brada{\v{c}}}, G.~{Brammer}, G.~{Desprez} et~al., \emph{{Bursty star formation and galaxy-galaxy interactions in low-mass galaxies 1 Gyr after the Big Bang}}, \href{https://doi.org/10.1093/MNRAS/stad3902}{\emph{Mon. Not. Roy. Astron. Soc.} {\bfseries 527} (2024) 11372} [\href{https://arxiv.org/abs/2310.02314}{{\ttfamily 2310.02314}}].

\bibitem{Gelli2023}
V.~{Gelli}, S.~{Salvadori}, A.~{Ferrara}, A.~{Pallottini} and S.~{Carniani}, \emph{{Quiescent Low-mass Galaxies Observed by JWST in the Epoch of Reionization}}, \href{https://doi.org/10.3847/2041-8213/acee80}{\emph{Astrophys. J. Lett.} {\bfseries 954} (2023) L11} [\href{https://arxiv.org/abs/2303.13574}{{\ttfamily 2303.13574}}].

\bibitem{Dome2024}
T.~{Dome}, S.~{Tacchella}, A.~{Fialkov}, D.~{Ceverino}, A.~{Dekel}, O.~{Ginzburg} et~al., \emph{{Mini-quenching of z = 4-8 galaxies by bursty star formation}}, \href{https://doi.org/10.1093/MNRAS/stad3239}{\emph{Mon. Not. Roy. Astron. Soc.} {\bfseries 527} (2024) 2139} [\href{https://arxiv.org/abs/2305.07066}{{\ttfamily 2305.07066}}].

\bibitem{FaisstMorishita2024}
A.L.~{Faisst} and T.~{Morishita}, \emph{{Dead or Alive? How Bursty Star Formation and Patchy Dust Can Cause Temporary Quiescence in High-redshift Galaxies}}, \href{https://doi.org/10.3847/1538-4357/ad58e2}{\emph{Astrophys. J.} {\bfseries 971} (2024) 47} [\href{https://arxiv.org/abs/2402.13316}{{\ttfamily 2402.13316}}].

\bibitem{Looser2024}
T.J.~{Looser}, F.~{D'Eugenio}, R.~{Maiolino}, J.~{Witstok}, L.~{Sandles}, E.~{Curtis-Lake} et~al., \emph{{A recently quenched galaxy 700 million years after the Big Bang}}, \href{https://doi.org/10.1038/s41586-024-07227-0}{\emph{Nature} {\bfseries 629} (2024) 53} [\href{https://arxiv.org/abs/2302.14155}{{\ttfamily 2302.14155}}].

\bibitem{Weisz2012}
D.R.~{Weisz}, B.D.~{Johnson}, L.C.~{Johnson}, E.D.~{Skillman}, J.C.~{Lee}, R.C.~{Kennicutt} et~al., \emph{{Modeling the Effects of Star Formation Histories on H{\ensuremath{\alpha}} and Ultraviolet Fluxes in nearby Dwarf Galaxies}}, \href{https://doi.org/10.1088/0004-637X/744/1/44}{\emph{Astrophys. J.} {\bfseries 744} (2012) 44} [\href{https://arxiv.org/abs/1109.2905}{{\ttfamily 1109.2905}}].

\bibitem{Emami2019}
N.~{Emami}, B.~{Siana}, D.R.~{Weisz}, B.D.~{Johnson}, X.~{Ma} and K.~{El-Badry}, \emph{{A Closer Look at Bursty Star Formation with L$_{H{\ensuremath{\alpha}} }$ and L$_{UV}$ Distributions}}, \href{https://doi.org/10.3847/1538-4357/ab211a}{\emph{Astrophys. J.} {\bfseries 881} (2019) 71} [\href{https://arxiv.org/abs/1809.06380}{{\ttfamily 1809.06380}}].

\bibitem{Carnall2018}
A.C.~{Carnall}, R.J.~{McLure}, J.S.~{Dunlop} and R.~{Dav{\'e}}, \emph{{Inferring the star formation histories of massive quiescent galaxies with BAGPIPES: evidence for multiple quenching mechanisms}}, \href{https://doi.org/10.1093/MNRAS/sty2169}{\emph{Mon. Not. Roy. Astron. Soc.} {\bfseries 480} (2018) 4379} [\href{https://arxiv.org/abs/1712.04452}{{\ttfamily 1712.04452}}].

\bibitem{Boquien2019}
M.~{Boquien}, D.~{Burgarella}, Y.~{Roehlly}, V.~{Buat}, L.~{Ciesla}, D.~{Corre} et~al., \emph{{CIGALE: a python Code Investigating GALaxy Emission}}, \href{https://doi.org/10.1051/0004-6361/201834156}{\emph{Astron. Astrophys.} {\bfseries 622} (2019) A103} [\href{https://arxiv.org/abs/1811.03094}{{\ttfamily 1811.03094}}].

\bibitem{Johnson2021}
B.D.~{Johnson}, J.~{Leja}, C.~{Conroy} and J.S.~{Speagle}, \emph{{Stellar Population Inference with Prospector}}, \href{https://doi.org/10.3847/1538-4365/abef67}{\emph{Astrophys. J. Suppl.} {\bfseries 254} (2021) 22} [\href{https://arxiv.org/abs/2012.01426}{{\ttfamily 2012.01426}}].

\bibitem{Cappellari2023}
M.~{Cappellari}, \emph{{Full spectrum fitting with photometry in PPXF: stellar population versus dynamical masses, non-parametric star formation history and metallicity for 3200 LEGA-C galaxies at redshift z {\ensuremath{\approx}} 0.8}}, \href{https://doi.org/10.1093/MNRAS/stad2597}{\emph{Mon. Not. Roy. Astron. Soc.} {\bfseries 526} (2023) 3273} [\href{https://arxiv.org/abs/2208.14974}{{\ttfamily 2208.14974}}].

\bibitem{Broussard2019}
A.~{Broussard}, E.~{Gawiser}, K.~{Iyer}, P.~{Kurczynski}, R.S.~{Somerville}, R.~{Dav{\'e}} et~al., \emph{{Star Formation Stochasticity Measured from the Distribution of Burst Indicators}}, \href{https://doi.org/10.3847/1538-4357/ab04ad}{\emph{Astrophys. J.} {\bfseries 873} (2019) 74} [\href{https://arxiv.org/abs/1901.01192}{{\ttfamily 1901.01192}}].

\bibitem{CT2019}
N.~{Caplar} and S.~{Tacchella}, \emph{{Stochastic modelling of star-formation histories I: the scatter of the star-forming main sequence}}, \href{https://doi.org/10.1093/MNRAS/stz1449}{\emph{Mon. Not. Roy. Astron. Soc.} {\bfseries 487} (2019) 3845} [\href{https://arxiv.org/abs/1901.07556}{{\ttfamily 1901.07556}}].

\bibitem{Dominguez2015}
A.~{Dom{\'\i}nguez}, B.~{Siana}, A.M.~{Brooks}, C.R.~{Christensen}, G.~{Bruzual}, D.P.~{Stark} et~al., \emph{{Consequences of bursty star formation on galaxy observables at high redshifts}}, \href{https://doi.org/10.1093/MNRAS/stv1001}{\emph{Mon. Not. Roy. Astron. Soc.} {\bfseries 451} (2015) 839} [\href{https://arxiv.org/abs/1408.5788}{{\ttfamily 1408.5788}}].

\bibitem{Sparre2017}
M.~{Sparre}, C.C.~{Hayward}, R.~{Feldmann}, C.-A.~{Faucher-Gigu{\`e}re}, A.L.~{Muratov}, D.~{Kere{\v{s}}} et~al., \emph{{(Star)bursts of FIRE: observational signatures of bursty star formation in galaxies}}, \href{https://doi.org/10.1093/MNRAS/stw3011}{\emph{Mon. Not. Roy. Astron. Soc.} {\bfseries 466} (2017) 88} [\href{https://arxiv.org/abs/1510.03869}{{\ttfamily 1510.03869}}].

\bibitem{Faisst2019}
A.L.~{Faisst}, P.L.~{Capak}, N.~{Emami}, S.~{Tacchella} and K.L.~{Larson}, \emph{{The Recent Burstiness of Star Formation in Galaxies at z {\ensuremath{\sim}} 4.5 from H{\ensuremath{\alpha}} Measurements}}, \href{https://doi.org/10.3847/1538-4357/ab425b}{\emph{Astrophys. J.} {\bfseries 884} (2019) 133} [\href{https://arxiv.org/abs/1909.03076}{{\ttfamily 1909.03076}}].

\bibitem{JFV2021}
J.A.~{Flores Vel{\'a}zquez}, A.B.~{Gurvich}, C.-A.~{Faucher-Gigu{\`e}re}, J.S.~{Bullock}, T.K.~{Starkenburg}, J.~{Moreno} et~al., \emph{{The time-scales probed by star formation rate indicators for realistic, bursty star formation histories from the FIRE simulations}}, \href{https://doi.org/10.1093/MNRAS/staa3893}{\emph{Mon. Not. Roy. Astron. Soc.} {\bfseries 501} (2021) 4812} [\href{https://arxiv.org/abs/2008.08582}{{\ttfamily 2008.08582}}].

\bibitem{Atek2022}
H.~{Atek}, L.J.~{Furtak}, P.~{Oesch}, P.~{van Dokkum}, N.~{Reddy}, T.~{Contini} et~al., \emph{{The star formation burstiness and ionizing efficiency of low-mass galaxies}}, \href{https://doi.org/10.1093/MNRAS/stac360}{\emph{Mon. Not. Roy. Astron. Soc.} {\bfseries 511} (2022) 4464} [\href{https://arxiv.org/abs/2202.04081}{{\ttfamily 2202.04081}}].

\bibitem{Tacchella2022}
S.~{Tacchella}, A.~{Smith}, R.~{Kannan}, F.~{Marinacci}, L.~{Hernquist}, M.~{Vogelsberger} et~al., \emph{{H {\ensuremath{\alpha}} emission in local galaxies: star formation, time variability, and the diffuse ionized gas}}, \href{https://doi.org/10.1093/mnras/stac818}{\emph{Mon. Not. Roy. Astron. Soc.} {\bfseries 513} (2022) 2904} [\href{https://arxiv.org/abs/2112.00027}{{\ttfamily 2112.00027}}].

\bibitem{Sun2023}
G.~{Sun}, A.~{Lidz}, A.L.~{Faisst} and C.-A.~{Faucher-Gigu{\`e}re}, \emph{{Probing bursty star formation by cross-correlating extragalactic background light and galaxy surveys}}, \href{https://doi.org/10.1093/MNRAS/stad2000}{\emph{Mon. Not. Roy. Astron. Soc.} {\bfseries 524} (2023) 2395} [\href{https://arxiv.org/abs/2305.08847}{{\ttfamily 2305.08847}}].

\bibitem{Clarke2024}
L.~{Clarke}, A.E.~{Shapley}, R.L.~{Sanders}, M.W.~{Topping}, G.B.~{Brammer}, T.~{Bento} et~al., \emph{{The Star-Forming Main Sequence in JADES and CEERS at $z>1.4$: Investigating the Burstiness of Star Formation}}, \href{https://doi.org/10.48550/arXiv.2406.05178}{\emph{arXiv e-prints} (2024) arXiv:2406.05178} [\href{https://arxiv.org/abs/2406.05178}{{\ttfamily 2406.05178}}].

\bibitem{Carnall2019}
A.C.~{Carnall}, J.~{Leja}, B.D.~{Johnson}, R.J.~{McLure}, J.S.~{Dunlop} and C.~{Conroy}, \emph{{How to Measure Galaxy Star Formation Histories. I. Parametric Models}}, \href{https://doi.org/10.3847/1538-4357/ab04a2}{\emph{Astrophys. J.} {\bfseries 873} (2019) 44} [\href{https://arxiv.org/abs/1811.03635}{{\ttfamily 1811.03635}}].

\bibitem{Alarcon2023}
A.~{Alarcon}, A.P.~{Hearin}, M.R.~{Becker} and J.~{Chaves-Montero}, \emph{{Diffstar: a fully parametric physical model for galaxy assembly history}}, \href{https://doi.org/10.1093/MNRAS/stac3118}{\emph{Mon. Not. Roy. Astron. Soc.} {\bfseries 518} (2023) 562} [\href{https://arxiv.org/abs/2205.04273}{{\ttfamily 2205.04273}}].

\bibitem{Dekel2023}
A.~{Dekel}, K.C.~{Sarkar}, Y.~{Birnboim}, N.~{Mandelker} and Z.~{Li}, \emph{{Efficient formation of massive galaxies at cosmic dawn by feedback-free starbursts}}, \href{https://doi.org/10.1093/MNRAS/stad1557}{\emph{Mon. Not. Roy. Astron. Soc.} {\bfseries 523} (2023) 3201} [\href{https://arxiv.org/abs/2303.04827}{{\ttfamily 2303.04827}}].

\bibitem{Harikane2023}
Y.~{Harikane}, M.~{Ouchi}, M.~{Oguri}, Y.~{Ono}, K.~{Nakajima}, Y.~{Isobe} et~al., \emph{{A Comprehensive Study of Galaxies at z 9-16 Found in the Early JWST Data: Ultraviolet Luminosity Functions and Cosmic Star Formation History at the Pre-reionization Epoch}}, \href{https://doi.org/10.3847/1538-4365/acaaa9}{\emph{Astrophys. J. Suppl.} {\bfseries 265} (2023) 5} [\href{https://arxiv.org/abs/2208.01612}{{\ttfamily 2208.01612}}].

\bibitem{Mason2023}
C.A.~{Mason}, M.~{Trenti} and T.~{Treu}, \emph{{The brightest galaxies at cosmic dawn}}, \href{https://doi.org/10.1093/MNRAS/stad035}{\emph{Mon. Not. Roy. Astron. Soc.} {\bfseries 521} (2023) 497} [\href{https://arxiv.org/abs/2207.14808}{{\ttfamily 2207.14808}}].

\bibitem{Mirocha2023}
J.~{Mirocha} and S.R.~{Furlanetto}, \emph{{Balancing the efficiency and stochasticity of star formation with dust extinction in z {\ensuremath{\gtrsim}} 10 galaxies observed by JWST}}, \href{https://doi.org/10.1093/MNRAS/stac3578}{\emph{Mon. Not. Roy. Astron. Soc.} {\bfseries 519} (2023) 843} [\href{https://arxiv.org/abs/2208.12826}{{\ttfamily 2208.12826}}].

\bibitem{Shen2023}
X.~{Shen}, M.~{Vogelsberger}, M.~{Boylan-Kolchin}, S.~{Tacchella} and R.~{Kannan}, \emph{{The impact of UV variability on the abundance of bright galaxies at z {\ensuremath{\geq}} 9}}, \href{https://doi.org/10.1093/MNRAS/stad2508}{\emph{Mon. Not. Roy. Astron. Soc.} {\bfseries 525} (2023) 3254} [\href{https://arxiv.org/abs/2305.05679}{{\ttfamily 2305.05679}}].

\bibitem{Sun2023b}
G.~{Sun}, C.-A.~{Faucher-Gigu{\`e}re}, C.C.~{Hayward}, X.~{Shen}, A.~{Wetzel} and R.K.~{Cochrane}, \emph{{Bursty Star Formation Naturally Explains the Abundance of Bright Galaxies at Cosmic Dawn}}, \href{https://doi.org/10.3847/2041-8213/acf85a}{\emph{Astrophys. J. Lett.} {\bfseries 955} (2023) L35} [\href{https://arxiv.org/abs/2307.15305}{{\ttfamily 2307.15305}}].

\bibitem{KB2024}
A.~{Kravtsov} and V.~{Belokurov}, \emph{{Stochastic star formation and the abundance of $z>10$ UV-bright galaxies}}, \href{https://doi.org/10.48550/arXiv.2405.04578}{\emph{arXiv e-prints} (2024) arXiv:2405.04578} [\href{https://arxiv.org/abs/2405.04578}{{\ttfamily 2405.04578}}].

\bibitem{Nikolic2024}
I.~{Nikoli{\'c}}, A.~{Mesinger}, J.E.~{Davies} and D.~{Prelogovi{\'c}}, \emph{{The importance of stochasticity in determining galaxy emissivities and UV LFs during cosmic dawn and reionization}}, \href{https://doi.org/10.1051/0004-6361/202451213}{\emph{Astron. Astrophys.} {\bfseries 692} (2024) A142} [\href{https://arxiv.org/abs/2406.15237}{{\ttfamily 2406.15237}}].

\bibitem{Munoz2023}
J.B.~{Mu{\~n}oz}, J.~{Mirocha}, S.~{Furlanetto} and N.~{Sabti}, \emph{{Breaking degeneracies in the first galaxies with clustering}}, \href{https://doi.org/10.1093/MNRASl/slad115}{\emph{Mon. Not. Roy. Astron. Soc.} {\bfseries 526} (2023) L47} [\href{https://arxiv.org/abs/2306.09403}{{\ttfamily 2306.09403}}].

\bibitem{ChakrabortyChoudhury2024}
A.~{Chakraborty} and T.R.~{Choudhury}, \emph{{Modelling the star-formation activity and ionizing properties of high-redshift galaxies}}, \href{https://doi.org/10.1088/1475-7516/2024/07/078}{\emph{JCAP} {\bfseries 2024} (2024) 078} [\href{https://arxiv.org/abs/2404.02879}{{\ttfamily 2404.02879}}].

\bibitem{Gelli2024}
V.~{Gelli}, C.~{Mason} and C.C.~{Hayward}, \emph{{The impact of mass-dependent stochasticity at cosmic dawn}}, \href{https://doi.org/10.48550/arXiv.2405.13108}{\emph{arXiv e-prints} (2024) arXiv:2405.13108} [\href{https://arxiv.org/abs/2405.13108}{{\ttfamily 2405.13108}}].

\bibitem{Iyer2020}
K.G.~{Iyer}, S.~{Tacchella}, S.~{Genel}, C.C.~{Hayward}, L.~{Hernquist}, A.M.~{Brooks} et~al., \emph{{The diversity and variability of star formation histories in models of galaxy evolution}}, \href{https://doi.org/10.1093/MNRAS/staa2150}{\emph{Mon. Not. Roy. Astron. Soc.} {\bfseries 498} (2020) 430} [\href{https://arxiv.org/abs/2007.07916}{{\ttfamily 2007.07916}}].

\bibitem{Tacchella2020}
S.~{Tacchella}, J.C.~{Forbes} and N.~{Caplar}, \emph{{Stochastic modelling of star-formation histories II: star-formation variability from molecular clouds and gas inflow}}, \href{https://doi.org/10.1093/MNRAS/staa1838}{\emph{Mon. Not. Roy. Astron. Soc.} {\bfseries 497} (2020) 698} [\href{https://arxiv.org/abs/2006.09382}{{\ttfamily 2006.09382}}].

\bibitem{PK2023}
Y.~{Pan} and A.~{Kravtsov}, \emph{{Modelling Stochastic Star Formation History of Dwarf Galaxies in GRUMPY}}, \href{https://doi.org/10.48550/arXiv.2310.08636}{\emph{arXiv e-prints} (2023) arXiv:2310.08636} [\href{https://arxiv.org/abs/2310.08636}{{\ttfamily 2310.08636}}].

\bibitem{PF2023}
A.~{Pallottini} and A.~{Ferrara}, \emph{{Stochastic star formation in early galaxies: Implications for the James Webb Space Telescope}}, \href{https://doi.org/10.1051/0004-6361/202347384}{\emph{Astron. Astrophys.} {\bfseries 677} (2023) L4} [\href{https://arxiv.org/abs/2307.03219}{{\ttfamily 2307.03219}}].

\bibitem{Dome2025}
T.~{Dome}, S.~{Martin-Alvarez}, S.~{Tacchella}, Y.~{Yuan} and D.~{Sijacki}, \emph{{Increased burstiness at high redshift in multiphysics models combining supernova feedback, radiative transfer, and cosmic rays}}, \href{https://doi.org/10.1093/mnras/staf006}{\emph{Mon. Not. Roy. Astron. Soc.} {\bfseries 537} (2025) 629} [\href{https://arxiv.org/abs/2410.00113}{{\ttfamily 2410.00113}}].

\bibitem{Planck2016}
{Planck Collaboration}, P.A.R.~{Ade}, N.~{Aghanim}, M.~{Arnaud}, M.~{Ashdown}, J.~{Aumont} et~al., \emph{{Planck 2015 results. XIII. Cosmological parameters}}, \href{https://doi.org/10.1051/0004-6361/201525830}{\emph{Astron. Astrophys.} {\bfseries 594} (2016) A13} [\href{https://arxiv.org/abs/1502.01589}{{\ttfamily 1502.01589}}].

\bibitem{CAFG2018}
C.-A.~{Faucher-Gigu{\`e}re}, \emph{{A model for the origin of bursty star formation in galaxies}}, \href{https://doi.org/10.1093/MNRAS/stx2595}{\emph{Mon. Not. Roy. Astron. Soc.} {\bfseries 473} (2018) 3717} [\href{https://arxiv.org/abs/1701.04824}{{\ttfamily 1701.04824}}].

\bibitem{FM2022}
S.R.~{Furlanetto} and J.~{Mirocha}, \emph{{Bursty star formation during the Cosmic Dawn driven by delayed stellar feedback}}, \href{https://doi.org/10.1093/MNRAS/stac310}{\emph{Mon. Not. Roy. Astron. Soc.} {\bfseries 511} (2022) 3895} [\href{https://arxiv.org/abs/2109.04488}{{\ttfamily 2109.04488}}].

\bibitem{MenonPower2024}
A.~{Menon} and C.~{Power}, \emph{{On bursty star formation during cosmological reionisation {\textendash} how does it influence the baryon mass content of dark matter halos?}}, \href{https://doi.org/10.1017/pasa.2024.39}{\emph{Publ. Astron. Soc. Austral.} {\bfseries 41} (2024) e049} [\href{https://arxiv.org/abs/2405.03211}{{\ttfamily 2405.03211}}].

\bibitem{Tacchella2018}
S.~{Tacchella}, S.~{Bose}, C.~{Conroy}, D.J.~{Eisenstein} and B.D.~{Johnson}, \emph{{A Redshift-independent Efficiency Model: Star Formation and Stellar Masses in Dark Matter Halos at z {\ensuremath{\gtrsim}} 4}}, \href{https://doi.org/10.3847/1538-4357/aae8e0}{\emph{Astrophys. J.} {\bfseries 868} (2018) 92} [\href{https://arxiv.org/abs/1806.03299}{{\ttfamily 1806.03299}}].

\bibitem{WechslerTinker2018}
R.H.~{Wechsler} and J.L.~{Tinker}, \emph{{The Connection Between Galaxies and Their Dark Matter Halos}}, \href{https://doi.org/10.1146/annurev-astro-081817-051756}{\emph{Ann. Rev. Astron. Astrophys.} {\bfseries 56} (2018) 435} [\href{https://arxiv.org/abs/1804.03097}{{\ttfamily 1804.03097}}].

\bibitem{MihosHernquist1994}
J.C.~{Mihos} and L.~{Hernquist}, \emph{{Triggering of Starbursts in Galaxies by Minor Mergers}}, \href{https://doi.org/10.1086/187299}{\emph{Astrophys. J. Lett.} {\bfseries 425} (1994) L13}.

\bibitem{McBride2009}
J.~{McBride}, O.~{Fakhouri} and C.-P.~{Ma}, \emph{{Mass accretion rates and histories of dark matter haloes}}, \href{https://doi.org/10.1111/j.1365-2966.2009.15329.x}{\emph{Mon. Not. Roy. Astron. Soc.} {\bfseries 398} (2009) 1858} [\href{https://arxiv.org/abs/0902.3659}{{\ttfamily 0902.3659}}].

\bibitem{Kelson2016}
D.D.~{Kelson}, A.J.~{Benson} and L.E.~{Abramson}, \emph{{On the Origin and Evolution of the Galaxy Stellar Mass Function}}, \href{https://doi.org/10.48550/arXiv.1610.06566}{\emph{arXiv e-prints} (2016) arXiv:1610.06566} [\href{https://arxiv.org/abs/1610.06566}{{\ttfamily 1610.06566}}].

\bibitem{Angles-Alcazar2017}
D.~{Angl{\'e}s-Alc{\'a}zar}, C.-A.~{Faucher-Gigu{\`e}re}, D.~{Kere{\v{s}}}, P.F.~{Hopkins}, E.~{Quataert} and N.~{Murray}, \emph{{The cosmic baryon cycle and galaxy mass assembly in the FIRE simulations}}, \href{https://doi.org/10.1093/MNRAS/stx1517}{\emph{Mon. Not. Roy. Astron. Soc.} {\bfseries 470} (2017) 4698} [\href{https://arxiv.org/abs/1610.08523}{{\ttfamily 1610.08523}}].

\bibitem{Wan2024}
J.T.~{Wan}, S.~{Tacchella}, B.D.~{Johnson}, K.G.~{Iyer}, J.S.~{Speagle} and R.~{Maiolino}, \emph{{Stochastic prior for non-parametric star-formation histories}}, \href{https://doi.org/10.1093/MNRAS/stae1734}{\emph{Mon. Not. Roy. Astron. Soc.} {\bfseries 532} (2024) 4002} [\href{https://arxiv.org/abs/2404.14494}{{\ttfamily 2404.14494}}].

\bibitem{SF2016}
G.~{Sun} and S.R.~{Furlanetto}, \emph{{Constraints on the star formation efficiency of galaxies during the epoch of reionization}}, \href{https://doi.org/10.1093/MNRAS/stw980}{\emph{Mon. Not. Roy. Astron. Soc.} {\bfseries 460} (2016) 417} [\href{https://arxiv.org/abs/1512.06219}{{\ttfamily 1512.06219}}].

\bibitem{Mirocha2017}
J.~{Mirocha}, S.R.~{Furlanetto} and G.~{Sun}, \emph{{The global 21-cm signal in the context of the high- z galaxy luminosity function}}, \href{https://doi.org/10.1093/MNRAS/stw2412}{\emph{Mon. Not. Roy. Astron. Soc.} {\bfseries 464} (2017) 1365} [\href{https://arxiv.org/abs/1607.00386}{{\ttfamily 1607.00386}}].

\bibitem{SL2024}
J.~{Sipple} and A.~{Lidz}, \emph{{The Star Formation Efficiency during Reionization as Inferred from the Hubble Frontier Fields}}, \href{https://doi.org/10.3847/1538-4357/ad06a7}{\emph{Astrophys. J.} {\bfseries 961} (2024) 50} [\href{https://arxiv.org/abs/2306.12087}{{\ttfamily 2306.12087}}].

\bibitem{Sun2021}
G.~{Sun}, T.C.~{Chang}, B.D.~{Uzgil}, J.J.~{Bock}, C.M.~{Bradford}, V.~{Butler} et~al., \emph{{Probing Cosmic Reionization and Molecular Gas Growth with TIME}}, \href{https://doi.org/10.3847/1538-4357/abfe62}{\emph{Astrophys. J.} {\bfseries 915} (2021) 33} [\href{https://arxiv.org/abs/2012.09160}{{\ttfamily 2012.09160}}].

\bibitem{Furlanetto2017}
S.R.~{Furlanetto}, J.~{Mirocha}, R.H.~{Mebane} and G.~{Sun}, \emph{{A minimalist feedback-regulated model for galaxy formation during the epoch of reionization}}, \href{https://doi.org/10.1093/MNRAS/stx2132}{\emph{Mon. Not. Roy. Astron. Soc.} {\bfseries 472} (2017) 1576} [\href{https://arxiv.org/abs/1611.01169}{{\ttfamily 1611.01169}}].

\bibitem{Feldmann2024}
R.~{Feldmann}, M.~{Boylan-Kolchin}, J.S.~{Bullock}, O.~{{\c{C}}atmabacak}, C.-A.~{Faucher-Gigu{\`e}re}, C.C.~{Hayward} et~al., \emph{{Elevated UV luminosity density at Cosmic Dawn explained by non-evolving, weakly-mass dependent star formation efficiency}}, \href{https://doi.org/10.48550/arXiv.2407.02674}{\emph{arXiv e-prints} (2024) arXiv:2407.02674} [\href{https://arxiv.org/abs/2407.02674}{{\ttfamily 2407.02674}}].

\bibitem{Kelly2014}
B.C.~{Kelly}, A.C.~{Becker}, M.~{Sobolewska}, A.~{Siemiginowska} and P.~{Uttley}, \emph{{Flexible and Scalable Methods for Quantifying Stochastic Variability in the Era of Massive Time-domain Astronomical Data Sets}}, \href{https://doi.org/10.1088/0004-637X/788/1/33}{\emph{Astrophys. J.} {\bfseries 788} (2014) 33} [\href{https://arxiv.org/abs/1402.5978}{{\ttfamily 1402.5978}}].

\bibitem{Kannan2021}
R.~{Kannan}, M.~{Vogelsberger}, F.~{Marinacci}, L.V.~{Sales}, P.~{Torrey} and L.~{Hernquist}, \emph{{Dust entrainment in galactic winds}}, \href{https://doi.org/10.1093/MNRAS/stab416}{\emph{Mon. Not. Roy. Astron. Soc.} {\bfseries 503} (2021) 336} [\href{https://arxiv.org/abs/2002.01933}{{\ttfamily 2002.01933}}].

\bibitem{Ferrara2024}
A.~{Ferrara}, \emph{{Super-early JWST galaxies, outflows, and Ly{\ensuremath{\alpha}} visibility in the Epoch of Reionization}}, \href{https://doi.org/10.1051/0004-6361/202348321}{\emph{Astron. Astrophys.} {\bfseries 684} (2024) A207} [\href{https://arxiv.org/abs/2310.12197}{{\ttfamily 2310.12197}}].

\bibitem{Sabti2022}
N.~{Sabti}, J.B.~{Mu{\~n}oz} and D.~{Blas}, \emph{{Galaxy luminosity function pipeline for cosmology and astrophysics}}, \href{https://doi.org/10.1103/PhysRevD.105.043518}{\emph{Phys. Rev. D} {\bfseries 105} (2022) 043518} [\href{https://arxiv.org/abs/2110.13168}{{\ttfamily 2110.13168}}].

\bibitem{ES2009BPASS}
J.J.~{Eldridge} and E.R.~{Stanway}, \emph{{Spectral population synthesis including massive binaries}}, \href{https://doi.org/10.1111/j.1365-2966.2009.15514.x}{\emph{Mon. Not. Roy. Astron. Soc.} {\bfseries 400} (2009) 1019} [\href{https://arxiv.org/abs/0908.1386}{{\ttfamily 0908.1386}}].

\bibitem{Ferland2017}
G.J.~{Ferland}, M.~{Chatzikos}, F.~{Guzm{\'a}n}, M.L.~{Lykins}, P.A.M.~{van Hoof}, R.J.R.~{Williams} et~al., \emph{{The 2017 Release Cloudy}}, \href{https://doi.org/10.48550/arXiv.1705.10877}{\emph{RMxAA} {\bfseries 53} (2017) 385} [\href{https://arxiv.org/abs/1705.10877}{{\ttfamily 1705.10877}}].

\bibitem{Chabrier2003}
G.~{Chabrier}, \emph{{Galactic Stellar and Substellar Initial Mass Function}}, \href{https://doi.org/10.1086/376392}{\emph{Publ. Astron. Soc. Pac.} {\bfseries 115} (2003) 763} [\href{https://arxiv.org/abs/astro-ph/0304382}{{\ttfamily astro-ph/0304382}}].

\bibitem{ST1999}
R.K.~{Sheth} and G.~{Tormen}, \emph{{Large-scale bias and the peak background split}}, \href{https://doi.org/10.1046/j.1365-8711.1999.02692.x}{\emph{Mon. Not. Roy. Astron. Soc.} {\bfseries 308} (1999) 119} [\href{https://arxiv.org/abs/astro-ph/9901122}{{\ttfamily astro-ph/9901122}}].

\bibitem{Hopkins2023}
P.F.~{Hopkins}, A.B.~{Gurvich}, X.~{Shen}, Z.~{Hafen}, M.Y.~{Grudi{\'c}}, S.~{Kurinchi-Vendhan} et~al., \emph{{What causes the formation of discs and end of bursty star formation?}}, \href{https://doi.org/10.1093/MNRAS/stad1902}{\emph{Mon. Not. Roy. Astron. Soc.} {\bfseries 525} (2023) 2241} [\href{https://arxiv.org/abs/2301.08263}{{\ttfamily 2301.08263}}].

\bibitem{TF2020}
A.C.~{Trapp} and S.R.~{Furlanetto}, \emph{{A flexible analytic model of cosmic variance in the first billion years}}, \href{https://doi.org/10.1093/MNRAS/staa2828}{\emph{Mon. Not. Roy. Astron. Soc.} {\bfseries 499} (2020) 2401} [\href{https://arxiv.org/abs/2009.05059}{{\ttfamily 2009.05059}}].

\bibitem{Knox1995}
L.~{Knox}, \emph{{Determination of inflationary observables by cosmic microwave background anisotropy experiments}}, \href{https://doi.org/10.1103/PhysRevD.52.4307}{\emph{Phys. Rev. D} {\bfseries 52} (1995) 4307} [\href{https://arxiv.org/abs/astro-ph/9504054}{{\ttfamily astro-ph/9504054}}].

\bibitem{Mehta2023}
V.~{Mehta}, H.I.~{Teplitz}, C.~{Scarlata}, X.~{Wang}, A.~{Alavi}, J.~{Colbert} et~al., \emph{{A Spatially Resolved Analysis of Star Formation Burstiness by Comparing UV and H{\ensuremath{\alpha}} in Galaxies at z {\ensuremath{\sim}} 1 with UVCANDELS}}, \href{https://doi.org/10.3847/1538-4357/acd9cf}{\emph{Astrophys. J.} {\bfseries 952} (2023) 133} [\href{https://arxiv.org/abs/2211.02056}{{\ttfamily 2211.02056}}].

\bibitem{Finkelstein2015}
S.L.~{Finkelstein}, J.~{Ryan}, Russell~E., C.~{Papovich}, M.~{Dickinson}, M.~{Song}, R.S.~{Somerville} et~al., \emph{{The Evolution of the Galaxy Rest-frame Ultraviolet Luminosity Function over the First Two Billion Years}}, \href{https://doi.org/10.1088/0004-637X/810/1/71}{\emph{Astrophys. J.} {\bfseries 810} (2015) 71} [\href{https://arxiv.org/abs/1410.5439}{{\ttfamily 1410.5439}}].

\bibitem{Bouwens2017}
R.J.~{Bouwens}, P.A.~{Oesch}, G.D.~{Illingworth}, R.S.~{Ellis} and M.~{Stefanon}, \emph{{The z {\ensuremath{\sim}} 6 Luminosity Function Fainter than -15 mag from the Hubble Frontier Fields: The Impact of Magnification Uncertainties}}, \href{https://doi.org/10.3847/1538-4357/aa70a4}{\emph{Astrophys. J.} {\bfseries 843} (2017) 129} [\href{https://arxiv.org/abs/1610.00283}{{\ttfamily 1610.00283}}].

\bibitem{Harikane2022}
Y.~{Harikane}, Y.~{Ono}, M.~{Ouchi}, C.~{Liu}, M.~{Sawicki}, T.~{Shibuya} et~al., \emph{{GOLDRUSH. IV. Luminosity Functions and Clustering Revealed with 4,000,000 Galaxies at z 2-7: Galaxy-AGN Transition, Star Formation Efficiency, and Implication for Evolution at z > 10}}, \href{https://doi.org/10.3847/1538-4365/ac3dfc}{\emph{Astrophys. J. Suppl.} {\bfseries 259} (2022) 20} [\href{https://arxiv.org/abs/2108.01090}{{\ttfamily 2108.01090}}].

\bibitem{Dalmasso2024}
N.~{Dalmasso}, M.~{Trenti} and N.~{Leethochawalit}, \emph{{Galaxy clustering measurements out to redshift z {\ensuremath{\sim}} 8 from Hubble Legacy Fields}}, \href{https://doi.org/10.1093/MNRAS/stad3901}{\emph{Mon. Not. Roy. Astron. Soc.} {\bfseries 528} (2024) 898} [\href{https://arxiv.org/abs/2312.12329}{{\ttfamily 2312.12329}}].

\bibitem{Sun2023HALF}
F.~{Sun}, E.~{Egami}, N.~{Pirzkal}, M.~{Rieke}, S.~{Baum}, M.~{Boyer} et~al., \emph{{First Sample of H{\ensuremath{\alpha}}+[O III]{\ensuremath{\lambda}}5007 Line Emitters at z > 6 Through JWST/NIRCam Slitless Spectroscopy: Physical Properties and Line-luminosity Functions}}, \href{https://doi.org/10.3847/1538-4357/acd53c}{\emph{Astrophys. J.} {\bfseries 953} (2023) 53} [\href{https://arxiv.org/abs/2209.03374}{{\ttfamily 2209.03374}}].

\bibitem{Covelo-Paz2024}
A.~{Covelo-Paz}, E.~{Giovinazzo}, P.A.~{Oesch}, R.A.~{Meyer}, A.~{Weibel}, G.~{Brammer} et~al., \emph{{An H-alpha view of galaxy build-up in the first 2 Gyr: luminosity functions at z\raisebox{-0.5ex}\textasciitilde4-6.5 from NIRCam/grism spectroscopy}}, \href{https://doi.org/10.48550/arXiv.2409.17241}{\emph{arXiv e-prints} (2024) arXiv:2409.17241} [\href{https://arxiv.org/abs/2409.17241}{{\ttfamily 2409.17241}}].

\bibitem{Casey2023}
C.M.~{Casey}, J.S.~{Kartaltepe}, N.E.~{Drakos}, M.~{Franco}, S.~{Harish}, L.~{Paquereau} et~al., \emph{{COSMOS-Web: An Overview of the JWST Cosmic Origins Survey}}, \href{https://doi.org/10.3847/1538-4357/acc2bc}{\emph{Astrophys. J.} {\bfseries 954} (2023) 31} [\href{https://arxiv.org/abs/2211.07865}{{\ttfamily 2211.07865}}].

\bibitem{Shivaei2015}
I.~{Shivaei}, N.A.~{Reddy}, A.E.~{Shapley}, M.~{Kriek}, B.~{Siana}, B.~{Mobasher} et~al., \emph{{The MOSDEF Survey: Dissecting the Star Formation Rate versus Stellar Mass Relation Using H{\ensuremath{\alpha}} and H{\ensuremath{\beta}} Emission Lines at z {\ensuremath{\sim}} 2}}, \href{https://doi.org/10.1088/0004-637X/815/2/98}{\emph{Astrophys. J.} {\bfseries 815} (2015) 98} [\href{https://arxiv.org/abs/1507.03017}{{\ttfamily 1507.03017}}].

\bibitem{Shapley2023}
A.E.~{Shapley}, R.L.~{Sanders}, N.A.~{Reddy}, M.W.~{Topping} and G.B.~{Brammer}, \emph{{JWST/NIRSpec Balmer-line Measurements of Star Formation and Dust Attenuation at z 3-6}}, \href{https://doi.org/10.3847/1538-4357/acea5a}{\emph{Astrophys. J.} {\bfseries 954} (2023) 157} [\href{https://arxiv.org/abs/2301.03241}{{\ttfamily 2301.03241}}].

\bibitem{Ferrara2024z14}
A.~{Ferrara}, \emph{{The eventful life of GS-z14-0, the most distant galaxy at redshift z = 14.32}}, \href{https://doi.org/10.1051/0004-6361/202450944}{\emph{Astron. Astrophys.} {\bfseries 689} (2024) A310} [\href{https://arxiv.org/abs/2405.20370}{{\ttfamily 2405.20370}}].

\bibitem{Katz2019}
H.~{Katz}, T.P.~{Galligan}, T.~{Kimm}, J.~{Rosdahl}, M.G.~{Haehnelt}, J.~{Blaizot} et~al., \emph{{Probing cosmic dawn with emission lines: predicting infrared and nebular line emission for ALMA and JWST}}, \href{https://doi.org/10.1093/mnras/stz1672}{\emph{Mon. Not. Roy. Astron. Soc.} {\bfseries 487} (2019) 5902} [\href{https://arxiv.org/abs/1901.01272}{{\ttfamily 1901.01272}}].

\bibitem{Kado-Fong2020}
E.~{Kado-Fong}, J.-G.~{Kim}, E.C.~{Ostriker} and C.-G.~{Kim}, \emph{{Diffuse Ionized Gas in Simulations of Multiphase, Star-forming Galactic Disks}}, \href{https://doi.org/10.3847/1538-4357/ab9abd}{\emph{Astrophys. J.} {\bfseries 897} (2020) 143} [\href{https://arxiv.org/abs/2006.06697}{{\ttfamily 2006.06697}}].

\bibitem{Wilkins2020}
S.M.~{Wilkins}, C.C.~{Lovell}, C.~{Fairhurst}, Y.~{Feng}, T.~{Di Matteo}, R.~{Croft} et~al., \emph{{Nebular-line emission during the Epoch of Reionization}}, \href{https://doi.org/10.1093/mnras/staa649}{\emph{Mon. Not. Roy. Astron. Soc.} {\bfseries 493} (2020) 6079} [\href{https://arxiv.org/abs/1904.07504}{{\ttfamily 1904.07504}}].

\bibitem{Eldridge2012}
J.J.~{Eldridge}, \emph{{Stochasticity, a variable stellar upper mass limit, binaries and star formation rate indicators}}, \href{https://doi.org/10.1111/j.1365-2966.2012.20662.x}{\emph{Mon. Not. Roy. Astron. Soc.} {\bfseries 422} (2012) 794} [\href{https://arxiv.org/abs/1106.4311}{{\ttfamily 1106.4311}}].

\bibitem{Peters2017}
T.~{Peters}, T.~{Naab}, S.~{Walch}, S.C.O.~{Glover}, P.~{Girichidis}, E.~{Pellegrini} et~al., \emph{{The SILCC project - IV. Impact of dissociating and ionizing radiation on the interstellar medium and H{\ensuremath{\alpha}} emission as a tracer of the star formation rate}}, \href{https://doi.org/10.1093/mnras/stw3216}{\emph{Mon. Not. Roy. Astron. Soc.} {\bfseries 466} (2017) 3293} [\href{https://arxiv.org/abs/1610.06569}{{\ttfamily 1610.06569}}].

\bibitem{Iyer2024}
K.G.~{Iyer}, J.S.~{Speagle}, N.~{Caplar}, J.C.~{Forbes}, E.~{Gawiser}, J.~{Leja} et~al., \emph{{Stochastic Modeling of Star Formation Histories. III. Constraints from Physically Motivated Gaussian Processes}}, \href{https://doi.org/10.3847/1538-4357/acff64}{\emph{Astrophys. J.} {\bfseries 961} (2024) 53} [\href{https://arxiv.org/abs/2208.05938}{{\ttfamily 2208.05938}}].

\bibitem{Schive2016}
H.-Y.~{Schive}, T.~{Chiueh}, T.~{Broadhurst} and K.-W.~{Huang}, \emph{{Contrasting Galaxy Formation from Quantum Wave Dark Matter, {\ensuremath{\psi}}DM, with {\ensuremath{\Lambda}}CDM, using Planck and Hubble Data}}, \href{https://doi.org/10.3847/0004-637X/818/1/89}{\emph{Astrophys. J.} {\bfseries 818} (2016) 89} [\href{https://arxiv.org/abs/1508.04621}{{\ttfamily 1508.04621}}].

\bibitem{Dayal2024}
P.~{Dayal} and S.K.~{Giri}, \emph{{Warm dark matter constraints from the JWST}}, \href{https://doi.org/10.1093/MNRAS/stae176}{\emph{Mon. Not. Roy. Astron. Soc.} {\bfseries 528} (2024) 2784} [\href{https://arxiv.org/abs/2303.14239}{{\ttfamily 2303.14239}}].

\bibitem{Sipple2024}
J.~{Sipple}, A.~{Lidz}, D.~{Grin} and G.~{Sun}, \emph{{Fuzzy Dark Matter Constraints from the Hubble Frontier Fields}}, \href{https://doi.org/10.48550/arXiv.2407.17059}{\emph{arXiv e-prints} (2024) arXiv:2407.17059} [\href{https://arxiv.org/abs/2407.17059}{{\ttfamily 2407.17059}}].

\end{thebibliography}\endgroup



\end{document}